\documentclass[a4paper,11pt]{article}

\usepackage{siunitx}
\usepackage{subfigure}
\usepackage{lineno}
\usepackage{float}

\pdfoutput = 1
\usepackage{jheppub} 

\usepackage[T1]{fontenc} 

\title{Calibration Strategy of the JUNO Experiment}

\collaboration{JUNO collaboration}
\author[5]{Angel Abusleme}
\author[45]{Thomas Adam}
\author[67]{Shakeel Ahmad}
\author[67]{Rizwan Ahmed}
\author[55]{Sebastiano Aiello}
\author[67]{Muhammad Akram}
\author[29]{Fengpeng An}
\author[10]{Guangpeng An}
\author[22]{Qi An}
\author[55]{Giuseppe Andronico}
% ...
\affiliation[5]{Pontificia Universidad Cat\'{o}lica de Chile, Santiago, Chile}
% ...
\affiliation[10]{Institute of High Energy Physics, Beijing, China}
% ...
\affiliation[22]{University of Science and Technology of China, Hefei, China}
% ...
\affiliation[29]{East China University of Science and Technology, Shanghai, China}
% ...
\affiliation[45]{IPHC, Universit\'{e} de Strasbourg, CNRS/IN2P3, F-67037 Strasbourg, France}
% ...
\affiliation[55]{INFN Catania and Dipartimento di Fisica e Astronomia dell Universit\`{a} di Catania, Catania, Italy}
% ...
\affiliation[67]{Pakistan Institute of Nuclear Science and Technology, Islamabad, Pakistan}
% ...

\emailAdd{Juno\_pub\_comm@juno.ihep.ac.cn}

\abstract{
  We present the calibration strategy for the 20 kton liquid
  scintillator central detector of the Jiangmen Underground Neutrino
  Observatory (JUNO). By utilizing a comprehensive multiple-source and
  multiple-positional calibration program, in combination with
  a novel dual calorimetry technique exploiting
  two independent photosensors and readout systems,
  we demonstrate that the JUNO central detector can achieve a
  better than 1\% energy linearity
  and a 3\% effective energy resolution,
  required by the neutrino mass ordering determination.}


\begin{document} 
\maketitle
\flushbottom

\section{Introduction}
\label{sec:intro}
Neutrino oscillation experiments~\cite{PDG}
have firmly established that there are three neutrino mass states with eigenvalues $m_1$, $m_2$, and $m_3$, 
and that at least two of these values are known to be different from zero. 
Solar neutrino experiments have established that
$m_1 < m_2$. 
However, 
there has been no clear experimental evidence whether $m_1<m_2<m_3$ or $m_3<m_1<m_2$, conventionally referred to as the normal or inverted neutrino mass ordering (MO), respectively.

The Jiangmen Underground Neutrino Observatory (JUNO) is a
multi-purpose experiment designed to elucidate fundamental neutrino properties, 
study neutrinos with astrophysical or terrestrial
origins, and search for rare processes beyond the Standard Model of particle physics~\cite{yellow-book}. 
Its 20 kton liquid scintillator (LS) central detector (CD) is located 680~m 
(\textcolor{black}{1775~m water equivalent}) underground in
Guangdong, China. The detector is at an equal distance of 53 km from
the Yangjiang and Taishan nuclear power plants.
Such a medium baseline configuration is ideal for the
determination of neutrino MO using the electron antineutrinos from
reactors~\cite{petcov-original-paper, petcov2, learned,zhanliang-original-paper-2008}. Due to neutrino oscillation, the
survival probability of electron antineutrinos can be written as:
\begin{eqnarray}
\label{eq_p_neutrino}
\begin{aligned}
  &P_{\overline{\nu}_{e} \rightarrow \overline{\nu}_{e}} = 1-\sin ^{2}2\theta_{13}\times \left( \cos^{2}\theta_{12} \sin^{2}\left(\Delta m_{31}^2\frac{L}{4E^{\nu}}\right) +\sin ^{2} \theta_{12} \sin ^{2}\left(\Delta m_{32}^2\frac{L}{4E^{\nu}}\right) \right) \\
  & -\cos ^{4} \theta_{13} \sin ^{2} 2 \theta_{12} \sin ^{2}\left(\Delta
  m_{21}^2\frac{L}{4E^{\nu}}\right)\,.
\end{aligned}
\end{eqnarray}
In this expression, $\theta_{12}$ and $\theta_{13}$ are the neutrino
mixing angles, $\Delta m_{ij}^2 \equiv m_i^2 - m_j^2$ is the
mass-square-difference between eigenstates $i$ and $j$, $L$ is the
distance from neutrino production to detection, and
$E^{\nu}$ is the \textcolor{black}{antineutrino} energy. 
Numerically, $\Delta m_{21}^2\sim7.5\times10^{-5}$~eV$^2$, and 
$|\Delta m_{32}^2|$ and $|\Delta m_{31}^2|$ are much larger ($\sim2.5\times10^{-3}$~eV$^2$). The normal or inverted MO corresponds to $|\Delta m_{31}^2| > |\Delta m_{32}^2|$ or 
$|\Delta m_{32}^2| > |\Delta m_{31}^2|$. 
Due to the fact that $\theta_{12}~\sim$~$\ang{34}$, 
the two oscillation frequencies driven by $\Delta m_{31}^2$ and
$\Delta m_{32}^2$ are weighted differently in Eq.~\eqref{eq_p_neutrino}.
This \textcolor{black}{allows} the determination of the MO
if the oscillation pattern in $E^{\nu}$ can be measured with
very high precision.

The JUNO LS is contained inside a 35-m diameter acrylic sphere with
12~cm thickness, strengthened by 591 stainless steel connection
bars. About 17,600 20-inch and \textcolor{black}{25,600}
3-inch photomultiplier \textcolor{black}{tubes} (PMTs)
are closely packed and immersed in ultra-pure water outside the \textcolor{black}{acrylic sphere oriented towards}
  the LS.
  These two sets of photosensors and their
  corresponding readout electronics constitute the LPMT (large) and
  SPMT (small) systems, forming the basis for the dual calorimetry discussed below.

  The electron antineutrinos are detected
via the so-called inverse $\beta$-decay (IBD), $\bar{\nu}_{e}+p
\rightarrow e^{+}+n$, where the electron antineutrinos interact with
protons in the LS to produce positrons and neutrons.
 The positron creates a prompt deposit of kinetic energy
 within a range from 0 to about 8 MeV in the LS,
 then annihilates with an electron producing gamma rays.
The energy of the anti-neutrino is related to the kinetic energy of
the positron as E$^{\nu}$ $\approx$ E$^{e+}+1.8$~MeV. The neutron will
be captured on hydrogen (99\%) or carbon (1\%) \textcolor{black}{within an average delay time of $\sim$200~$\mu$s,}
producing gammas of 2.22~MeV or 4.95~MeV, respectively.

Particle interactions in the LS will produce scintillating
(dominant) and Cherenkov (sub-dominant, $\leq$10\%) photons,
in numbers mostly proportional to the deposited energy,
which will then be converted to photoelectrons (PEs) by the PMTs and digitized
by the electronics.
However, there is an intrinsic non-linearity in both
light emitting mechanisms.
Scintillation photon yield follows a so-called Birks' law
  with ``quenching'' effect depending on the energy and
  type of particle~\cite{PDG}, whilst
  the Cherenkov emission depends on the
velocity of the charged particle. The combined effect will be referred to as the ``physics
non-linearity'' hereafter, which can be calibrated by the combination of
radioactive sources and natural radioactivity background. The PMT
  instrumentation and electronics may carry additional event-level
  ``instrumental non-linearity'', a nonlinear response between the
  created photons in the LS and the measured charge from the
  electronics, originated from channel-level instrumental
  non-linearity. For JUNO, this effect is particularly delicate, since
  at a given energy the charge response of single LPMT varies by more
  than two orders of magnitude.
  A novel methodology called the dual calorimetry is designed, 
  utilizing a comparison between the LPMT and the SPMT systems 
  which work under very different photon occupancy regimes, to directly 
  calibrate such non-linearity.

  In addition to the energy non-linearity, the total
  \textcolor{black}{number of PEs} collected by the LPMT and SPMT systems is also position-dependent
  in JUNO due to PMT solid angles, optical attenuation effects,
  reflections at material interfaces, shadowing due to opaque
  materials, etc. This intrinsic position
  non-uniformity, mostly energy independent, has to be corrected by a
  multi-positional calibration to optimize the energy resolution.

The determination of MO requires that the uncertainty of the positron
kinetic energy scale should be better than 1\%. Moreover, the effective
energy resolution has to be better than 3\%~\cite{zhanliang-original-paper-2009,zhanliang-original-paper-2013,yellow-book},
\textcolor{black}{an unprecedented energy resolution in any of the LS-based neutrino experiment.}
To achieve these stringent requirements,
not only the detector performance has to be \textcolor{black}{excellent} (e.g. high light yield and LS transparency),
but a comprehensive calibration program is also a must. Some early concepts of JUNO calibration have been introduced in Ref.~\cite{JUNOCDR}.
In this paper, however, we focus on the
calibration strategy for JUNO and demonstrate how it helps satisfy the physics requirements.

The remainder of the paper is organized as follows. In
section~{\ref{sec:scale_calib}}, we discuss the approach of the energy
scale calibration to tackle the physics and instrumental non-linearity.
We then develop the method to minimize the energy
resolution via correcting the position non-uniformity
in section~{\ref{sec:resolution}}. Based on these, we present the
conceptual design of the complete calibration hardware in
section~{\ref{sec:CalibSys}}, and the calibration program that we
envision in section~{\ref{sec:calib_program}}, before we conclude in
section~{\ref{sec:summary}}.

\section{Energy scale calibration}
\label{sec:scale_calib}
A custom Geant4-based (version 9.4.p04)~\cite{Geant4} software, called
SNIPER~\cite{SNIPER}, is used to perform \textcolor{black}{calibration-related}
simulations. \textcolor{black}{SNIPER contains among others} the up-to-date JUNO detector
geometry and optical parameters. The JUNO LS was
  tested in a decommissioned Daya Bay detector, so the LS optical
  parameters such as the light yield, absorption and re-emission
  probability are tuned to the data~\cite{YuZeyuan_LSpaper,
    JUNOLSAbsL}. The Rayleigh scattering length is obtained from a
  separate bench experiment~\cite{Zhouxiang_Paper}. The optical parameters of the
  acrylic sphere, ultrapure water and other materials are taken from
  bench measurements. The quantum efficiency and collection efficiency
  (angle-dependent) of the LPMTs are initially set at
  the average value from quality assurance tests and
  can be adjusted individually in the simulation.
 The ``low energy Livermore model'', which incorporates
 atomic shell cross section data~\cite{geant4-physics},
 is selected as the electromagnetic interaction model in SNIPER.

\subsection{Calibration of physics non-linearity}
\subsubsection{Selection of sources}

The radioactive sources and processes considered in JUNO and types of emitted
\textcolor{black}{radiation} are listed in table~\ref{radioactive_source}.
For a large LS detector, thin-walled electron or positron sources
would pose risk of leakage of radionuclides.
Instead, we consider $\gamma$ sources
ranging from a few hundred keV to a few MeV to cover the range of the
prompt energy of IBDs. \textcolor{black}{Concerning the $^{68}$Ge source, it decays in $^{68}$Ga
via electron capture, which then $\beta^{+}$-decays to $^{68}$Zn. The kinetic energy of the positrons will
be absorbed by the enclosure, so only the annihilation gammas are released.}
In addition, ($\alpha$,n) sources such as
$^{241}$Am-Be (AmBe) and $^{241}$Am-$^{13}$C (AmC) can be used to 
provide both high energy gammas and neutrons, the latter of which also produces capture gammas on
hydrogen and carbon atoms in the LS.

\begin{table}[H]
\centering
    \begin{tabular}{c|c|c}\hline
          Sources/Processes & Type & Radiation\\\hline
          $^{137}$Cs& $\gamma$ & 0.662 MeV\\
          $^{54}$Mn & $\gamma$ & 0.835 MeV\\
          $^{60}$Co & $\gamma$ & 1.173 + 1.333 MeV\\
          $^{40}$K  & $\gamma$ & 1.461 MeV\\
          $^{68}$Ge & e$^{+}$ & \textcolor{black}{annihilation} 0.511 + 0.511 MeV\\
          $^{241}$Am-Be & n, $\gamma$ & neutron + 4.43 MeV ($^{12}$C$^{*}$) \\
          $^{241}$Am-$^{13}$C &n, $\gamma$ & neutron + 6.13 MeV ($^{16}$O$^{*}$)\\
          (n,$\gamma$)p & $\gamma$ & 2.22 MeV \\
          (n,$\gamma$)$^{12}$C & $\gamma$ & 4.94 MeV or 3.68 + 1.26 MeV\\\hline
    \end{tabular}
  \caption{\label{radioactive_source}List of \textcolor{black}{radioactive sources and processes} considered in JUNO calibration. }
\end{table}

\subsubsection{Model of physics non-linearity}
\label{sec:nonlin}
Most IBD positrons lose energy by ionization
before they stop. A stopped positron can either
directly annihilate with an electron, or form para- or
ortho-positronium bound state before annihilation~\cite{borexino-positronium-paper,Ortho-positronium-double-chooz,measure-lifetime-ortho}.
The annihilation produces
either two or three gammas, respectively, with a total energy of 1.022 MeV.
We define a general term "visible energy" as the energy estimated based on the detected number of PEs 
\begin{eqnarray}
\label{eq:evis}
    E_{\rm vis} = \rm{PE}/Y_{0}\,,
\end{eqnarray}
where the light yield Y$_{0}$ is a constant obtained from calibration (see section~\ref{sec:calib_proc}).
The visible prompt energy of an IBD can then be decomposed as:
\begin{eqnarray}
\label{E_eplus_vis}
E_{\rm vis}^{\rm prompt} = E_{\rm vis}^{e} + E_{\rm vis}^{\rm anni}\,.
\end{eqnarray}
\textcolor{black}{The first term $E^{e}_{\rm vis}$ describes the visible energy associated with the kinetic energy of the positron,
which is approximately the same for an electron with the same energy.}
The second term $E_{\rm vis}^{\rm anni}$ is the visible
energy of the annihilation gammas, which can be
approximately calibrated using an enclosed $^{68}$Ge source.
The physics non-linearity is defined by:
\begin{eqnarray}
\label{E_nonlin}
f_{\rm nonlin} = \frac{E_{\rm vis}^e}{E^e}\,,
\end{eqnarray}
in which $E^e$ is the true kinetic energy of the electron or
positron. Therefore, if $f_{\rm nonlin}$ can be determined via calibration,
then for each IBD event, the true kinetic energy of a positron can be
reconstructed as:
\begin{eqnarray}
\label{eq:positron_energy_rec}
E^{e+}_{\rm{rec}} = \displaystyle\frac{E_{\rm vis}^{\rm prompt}-E_{\rm vis}^{\rm anni}}{f_{\rm nonlin}}\,.
\end{eqnarray}

A gamma deposits its energy into the LS via secondary electrons,
allowing a robust determination of $f_{\rm nonlin}$
using the gamma calibration
data~\cite{kamland-calibration-paper,dayabay-first-shape-paper,semi-ana}. Without
loss of generality, $f_{\rm nonlin}$ is modeled as a four parameter
function, as used in the first Daya Bay spectral
analysis~\cite{dayabay-first-shape-paper}:
\begin{eqnarray}
\label{Ele_nonlin}
f_{\rm nonlin} = \frac{p_{0}+p_{3}/E^e}{1+p_{1}e^{-p_{2}E^e}}\,, 
\end{eqnarray}
where the numerator is to include a first order non-linearity and
the denominator is to ensure a damping of the non-linearity at high energy,
as both Birks and Cherenkov contributions are expected to get more linear with energy.

The visible energy of the gamma can be written as
\begin{eqnarray}
\label{eq_gamma}
E_{\rm vis}^{\gamma} = \int^{E^{\gamma}}_{0} P(E^e)\times f_{\rm nonlin}(E^e) \times E^{e} dE^e \times \mathcal{I}\,.
\end{eqnarray}
In this expression, $P(E^e)$ denotes the probability density function of a given gamma source converted to secondary
electrons/positrons at an energy $E^{e}$ via Compton scattering, photoelectric effect, 
or pair production, determined from the simulation.
$\mathcal{I}$ is a normalization factor very close to unity 
\begin{eqnarray}
    \mathcal{I} = \frac{E^{\gamma}} {\int^{E^{\gamma}}_{0} P(E^e)\times E^{e} dE^e} 
\end{eqnarray}
to account for an order of 0.2\% missing energy in the simulation due to secondary production thresholds \textcolor{black}{of 250~eV}. 

The highest gamma
energy in table~\ref{radioactive_source} is 6.13~MeV, which is insufficient to cover the full
range of IBD positron energies. To extend the energy range in calibration, cosmogenically produced $^{12}$B, about 1000 events
per day in JUNO~\cite{yellow-book}, will be used. $^{12}$B decays via $\beta$-emissions with a $Q$ value of 13.4~MeV and a lifetime of 29~ms, with more than 98\%
into the ground state of $^{12}$C. Therefore it
offers complementary constraints to
$f_{\rm nonlin}(E^e)$ at the high energy end~\cite{dayabay-calib-paper}.
$^{12}\rm B$ events can be cleanly
  identified by looking for delayed high
  energy $\beta$ event (with a few percent mixture of $^{12}$N, a
  $\beta^{+}$-emitter) after an energetic
  muon~\cite{kamland-cosmogenic-bkg-paper,borexino-cosmogenic-bkg-paper}.

\subsubsection{Calibration procedure}
\label{sec:calib_proc}
\textcolor{black}{To validate our constraints to the physics non-linearity using calibration,
bare gamma sources in table~\ref{radioactive_source} are simulated from the center of the CD, while $^{12}$B}
decays are simulated uniformly in the detector. 
The light yield is determined to be Y$_{0}$=1345 PE/MeV by taking the simulated neutron captures on hydrogen at the 
CD center, and dividing the mean number of PEs by 2.22 MeV.
The visible energy of each event is then reconstructed by Eq.~\eqref{eq:evis}. 
100,000 events are simulated for each gamma source, for which the centroid is determined by Gaussian fit to a level of 0.01\% statistically. The statistics of $^{12}$B events is assumed to be equivalent to one month running period, \textcolor{black}{but inserting a 1~ms cut after their production after energetic muons to remove potential energy bias.}
The non-uniformity correction Eq.~\eqref{eq_E_rec} is applied to these events, as well as a 3~MeV threshold 
to suppress accidental background, \textcolor{black}{and the upper range 12~MeV covers the entire energy region of reactor IBDs.}
In addition, a radius cut of 15 m is chosen to avoid energy
leakage via Bremsstrahlung and complicated non-uniformity correction close to the boundary, 
leading to a 61\% acceptance.

To combine the gamma
sources and $^{12}$B data to fit for the four parameters of $f_{\rm nonlin}$, we define a $\chi^{2}$ as:
\begin{eqnarray}
\label{eq_chi2}
\textcolor{black}{\chi^{2}=\sum_{i=1}^{8}\left(\frac{M_{i}^{\gamma}-P_{i}^{\gamma}}{\sigma_{i}}\right)^{2}+\sum_{j=1}^{N} \frac{\left(M_{j}^{\rm B12}-P_{j}^{\rm B12}\right)^{2}}{M_{j}^{\rm B12}},}
\end{eqnarray}
where $M_{i}^{\gamma}$ and $P_{i}^{\gamma}$ are the measured and
predicted (see Eq.~\eqref{eq_gamma}) visible energy peaks of the $i$th gamma source, respectively,
and $\sigma_{i}$ is the statistical uncertainty of $M_{i}^{\gamma}$. The $^{12}$B
visible energy spectrum is binned into 90 bins from 3~MeV to 12~MeV, so $M_{j}^{\rm B12}$ and $P_{j}^{\rm B12}$
are the number of measured and predicted events in the $j$th bin.
In addition to $f_{\rm nonlin}$, $P_{j}^{\rm B12}$ also takes into account
spectral smearing according to
the energy resolution function Eq.~\eqref{eq_resolution}.
The systematic uncertainty of the calibration is neglected in Eq.~\eqref{eq_chi2},
but is included separately in the uncertainty band of
electron non-linearity using the procedure in section~{\ref{sec:sys_uncer}}.

\begin{figure}[H]
  \centering
  \subfigure[Gamma non-linearity]{
    \includegraphics[width=3in]{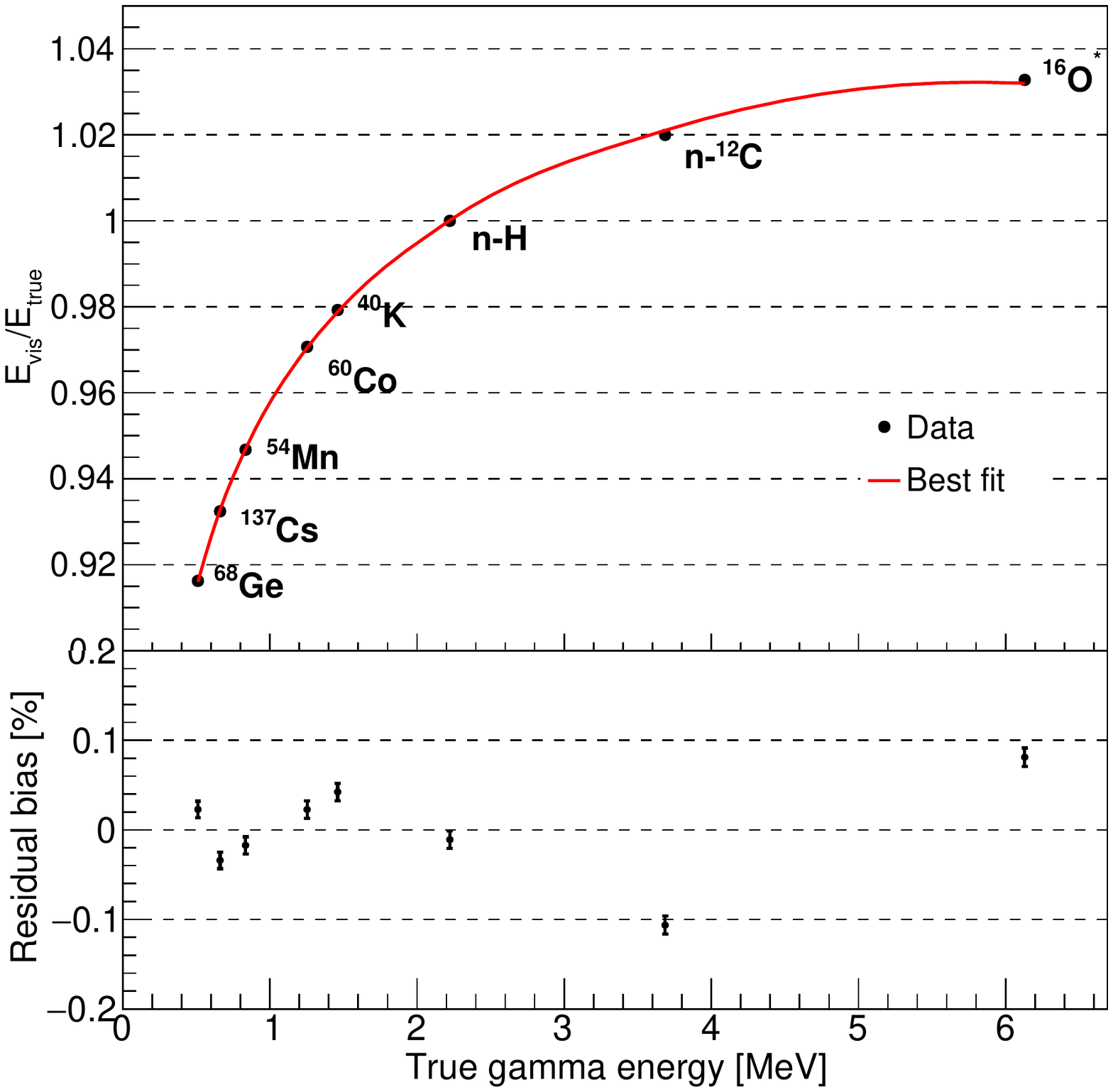}
    \label{non_linearity_fit}
  }
  \subfigure[Boron Spectrum]{
    \includegraphics[width=3in]{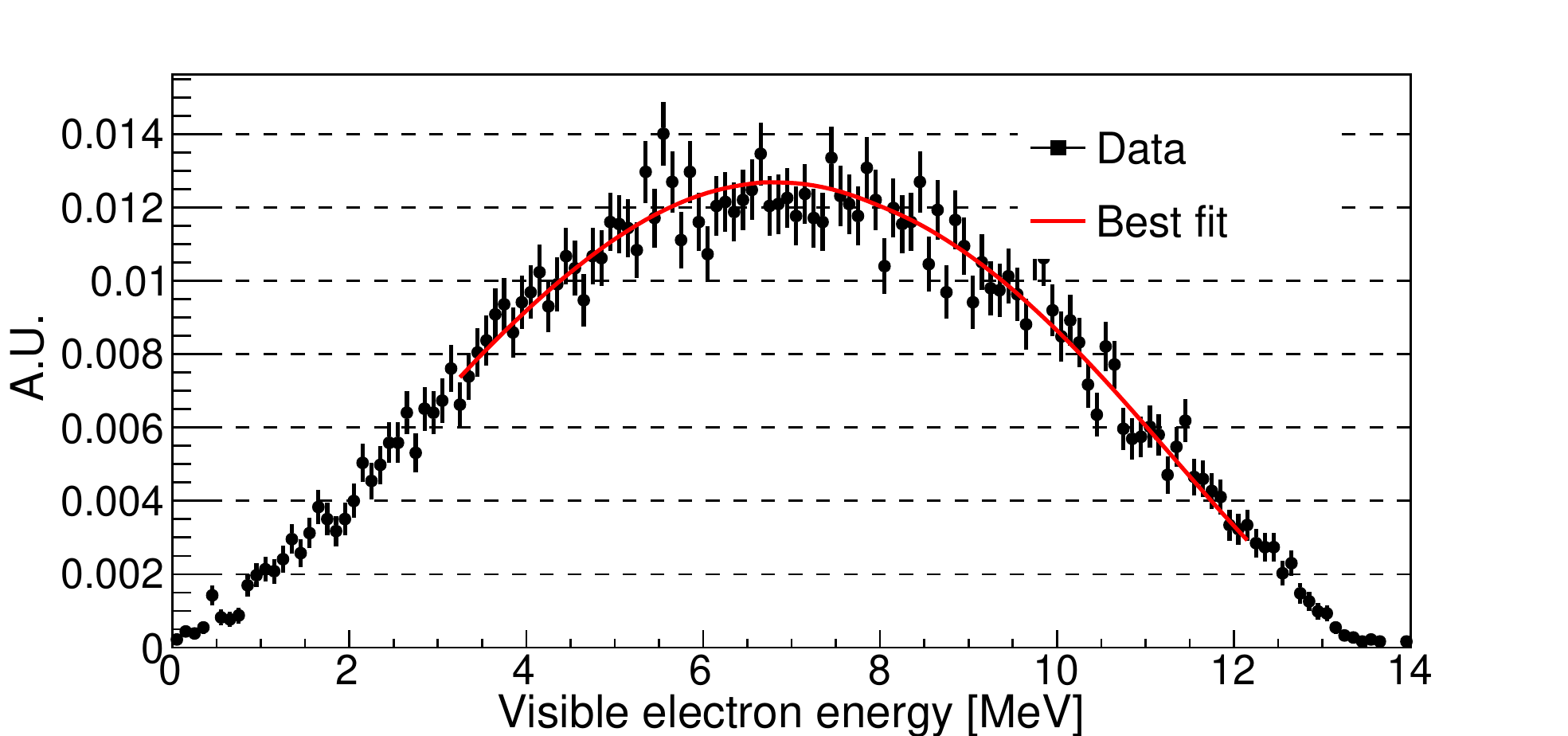}
    \label{B12}
  }
  \subfigure[Electron non-linearity]{
    \includegraphics[width=3in]{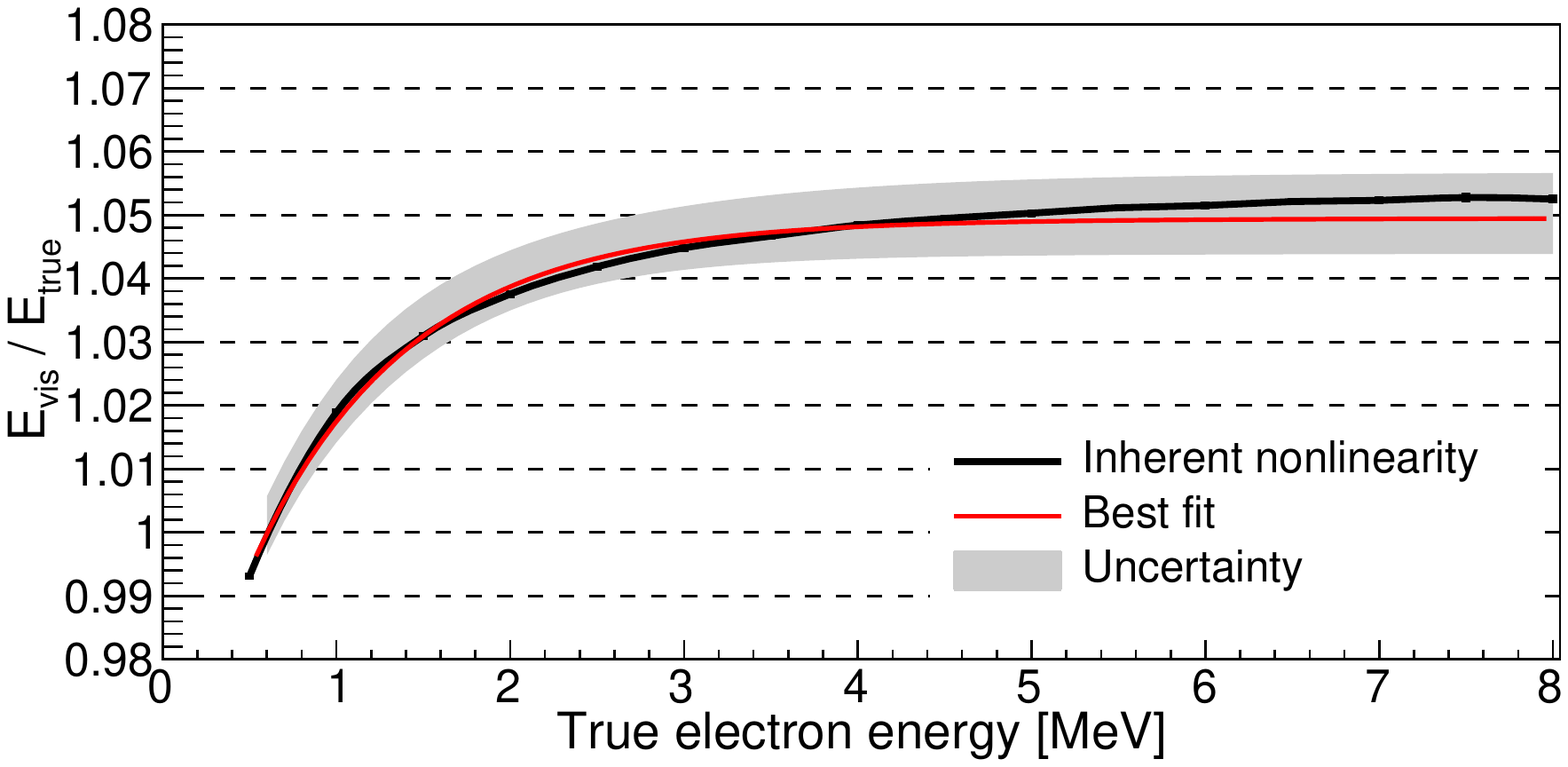}
    \label{non_e_linearity_fit}
  }
  \caption{Fitted and simulated gamma non-linearity (a), $^{12}$B
    spectra (b), and electron non-linearity (c). In all figures, black
    points (curves) represent simulated data, and red points
    (curves) are the best fits.
    For visual clarity,
    for sources with multiple gamma emissions, the
    horizontal axis is chosen to be the mean energy of the gammas.
    In (c), the uncertainty band is
    evaluated using the procedure in section~{\ref{sec:sys}}.}
\end{figure}

The non-linearity parameters $p_i$ in Eq.~\eqref{Ele_nonlin} can now be
determined using the $\chi^2$ fit. For the gamma sources, the
ratios of the best fit visible energy to the
true are compared to those from the simulated data in
figure~\ref{non_linearity_fit}. The best fit model for $^{12}$B spectrum is also overlaid with
simulated data in figure~\ref{B12}. Excellent agreements are observed in both figures.


For comparison, we use the \textcolor{black}{same SNIPER simulation} to produce the
visible energy for individual mono-energetic electrons at the CD
center and to extract the true inherent $f_{\rm nonlin}$.
\textcolor{black}{The best fit $f_{\rm nonlin}$
is displayed in figure~\ref{non_e_linearity_fit},
together with the true,
and they agree within 0.3\% within the entire energy range from 0.5 to 8 MeV.}

\begin{figure}
  \centering
  \includegraphics[width=3in]{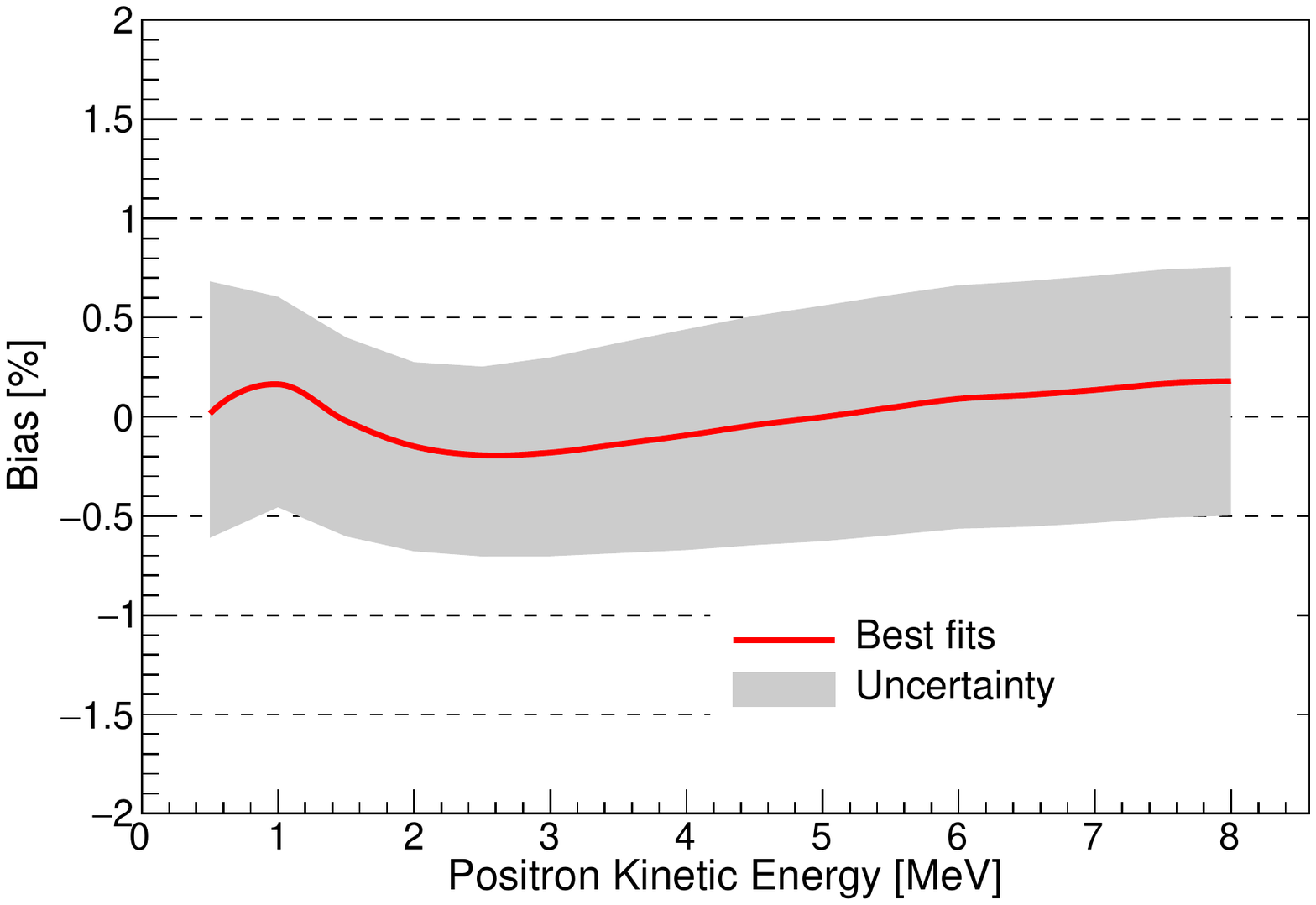}
  \caption{ 
  	\label{fig:positron_non-linearity} Bias $(E_{\rm {rec}}^{e+} - E^{e+})/E^{e+}$
    in the reconstructed positron kinetic energy as a
    function of true energy. 
	\textcolor{black}{The shape of the bias is an artifact that the
	degree-of-freedom of the fit function does not describe perfectly the true non-linearity curve.}
	The band represents the uncertainties in the calibration procedure,
    detailed in section~{\ref{sec:sys_uncer}}. 
	The lowest kinetic energy in the
    figure is 0.5~MeV, to avoid artificial increase of the fractional
    bias when the kinetic energy approaches zero.
	}
\end{figure}

For a final sanity check,
\textcolor{black}{a simulation of mono-energetic positrons at the CD center is performed.}
The kinetic energy is
reconstructed event-by-event using Eq.~\eqref{eq:positron_energy_rec},
in which $E_{\rm vis}^{\rm anni}$ is obtained from the simulation of
an enclosed $^{68}$Ge source, and the best fit $f_{\rm nonlin}$ is taken
from the previous step.
The residual bias in the reconstructed energy, as depicted in figure~\ref{fig:positron_non-linearity},
is within $\pm$ 0.2\%, with a 0.7\% uncertainty band to be discussed in section~{\ref{sec:sys_uncer}}.
The positron energy non-linearity could be further improved
by utilizing additional cosmogenic background in the detector, for example, $^{10}$C and $^{11}$C.

\subsection{Calibration of instrumental non-linearity}
\label{sec:elec_nonlin}
\textcolor{black}{As mentioned earlier, even at a given energy, the single channel response of LPMT is strongly position dependent in JUNO. Channel-level non-linearities between the actual photons and measured charge would convolve into an event-level instrumental non-linearity, which is consequently entangled with position non-uniformity.}

\textcolor{black}{To correct for this complication, 
the LPMT system is first calibrated for its channel-level
non-linearity utilizing the dual calorimetry calibration technique, which
implies the response comparison between the LPMT and SPMT calorimetries, with the help of a tunable light source covering the full range of the uniform IBDs (0 to 100~PE per LPMT channel). Such a UV laser system has been developed and described in Ref.~\cite{zhangyuanyuanpaper}, which can produce a uniform illumination on all channels over the desired range when flashing from the center of the CD. Within such range, the SPMT can serve as an approximate linear in-detector reference, ensured by both photon counting and charge measurement, since SPMT channels primarily operate in the single photon regime. As the laser intensity varies, the ratio of the LPMT charge to that of the total SPMTs leads to a direct determination of the LPMT channel-level non-linearity. 
This calibration scheme is immune from the physics non-linearity, since both LPMT and SPMT are exposed to the same energy
deposition. The non-uniformity is also irrelevant here, as the laser calibration source is kept in the detector center. 
}

\begin{figure}
  \centering
  \includegraphics[width=3.2in]{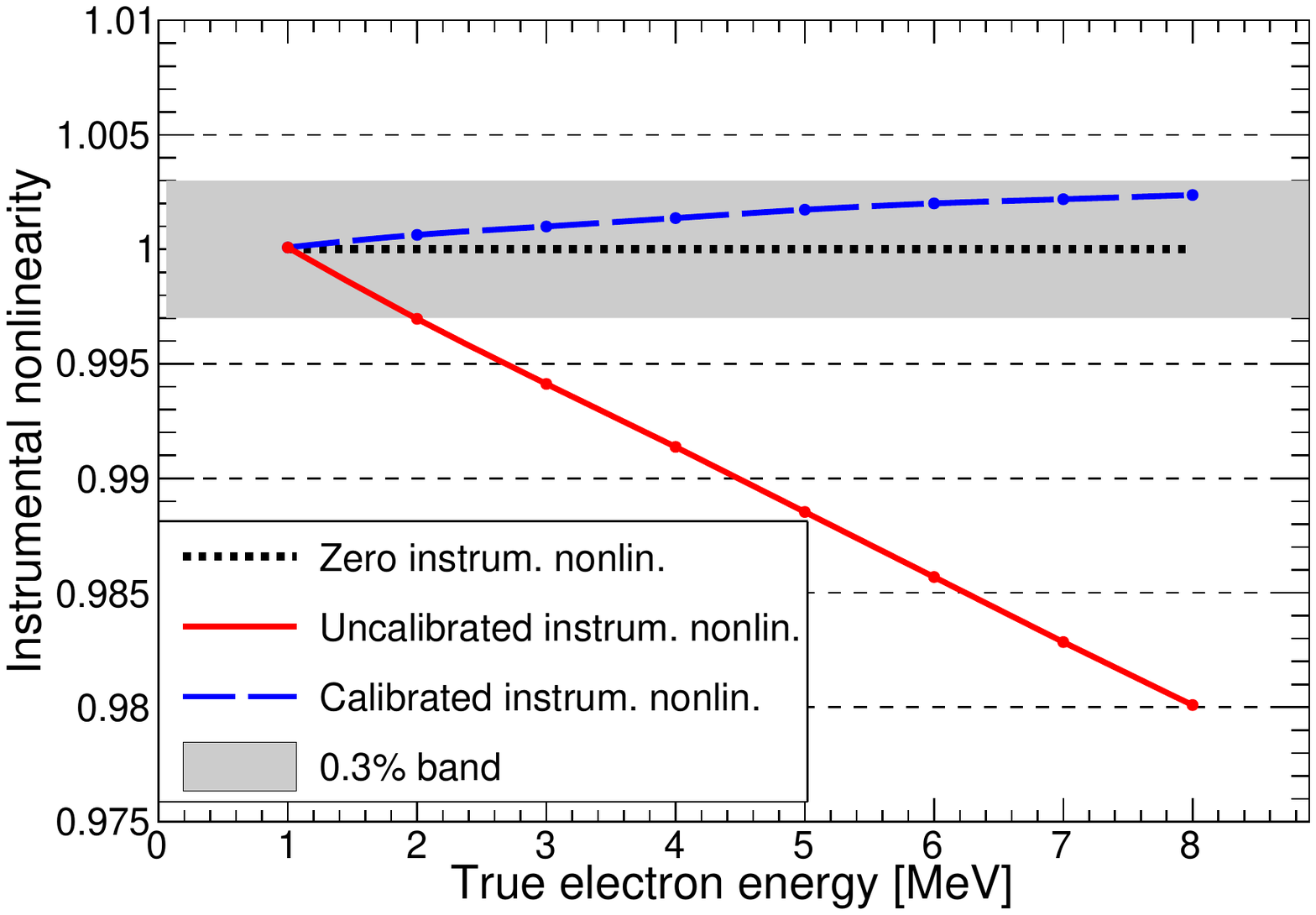}
  \caption{
  \label{fig:inst_non-linearity}
  \textcolor{black}{
  Event-level instrumental non-linearity,
  defined as the ratio of the total measured LPMT charge 
  to the true charge for events uniformly distributed in the detector. The dotted black line represents the perfect linear case.
 The solid red line represents event-level non-linearity without the channel-level correction, with position non-uniformity obtained at 1 MeV applied, in an extreme hypothetical scenario of 50\%
  non-linearity over 100 PEs for the LPMTs. The dashed blue line represents that after the channel-level correction. 
  The gray band shows the residual uncertainty of 0.3\%, after the channel-level correction.}
  }
\end{figure}

\textcolor{black}{To illustrate this approach, electron events from 1 to 8~MeV are simulated uniformly in the CD. An extreme channel-level non-linearity of 50\% over 100~PE for the LPMT is assumed.
As shown in figure~\ref{fig:inst_non-linearity}, 
the impact the channel-level instrumental non-linearity can lead
to an event-level non-linearity as large as 2\% at 8 MeV. 
For comparison, if the LPMT charge is first calibrated and corrected at the channel-level with the dual calorimetry approach, the residual event-level non-linearity is reduced to \textless 0.3\%.
}

\textcolor{black}{Residual biases could still remain after the laser calibration if the instrumental non-linearity depends on the photon arrival time profile, which may be different between the laser and physical events. This effect can be controlled by a systematic comparison of the LPMT charge responses between the laser and radioactive sources.
}

\subsection{Evaluation of systematic uncertainties}
\label{sec:sys}
The non-linearity calibration methodology above is designed upon the experience in Daya Bay~\cite{dayabay-calib-paper},
KamLAND~\cite{kamland-calib-paper}, Borexino~\cite{Borexino-calib-paper} and Double Chooz~\cite{DoubleChooz-Calib}. 
Residual systematic uncertainties and their combined effects to the positron energy scale are evaluated in this section.

\subsubsection{Shadowing effect}
\begin{figure}
  \centering
  \includegraphics[width=4.2in]{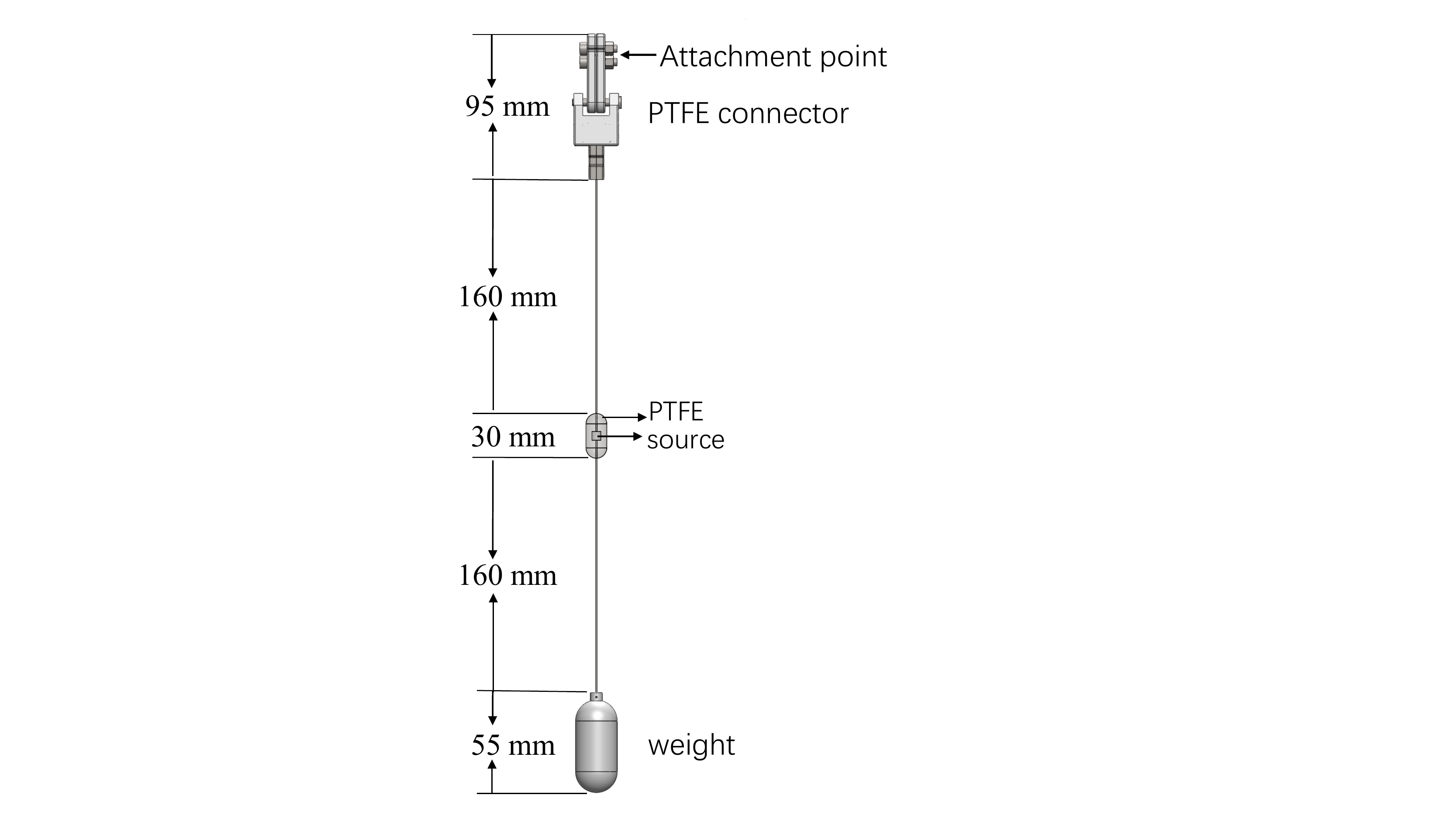}
  \caption{
  \label{source_weight_QC}
	Design of a typical source assembly.
  The $6\times6$~mm$^{2}$ stainless steel source capsule is enclosed in a 30 mm high PTFE shell.}
\end{figure}

A realistic \textcolor{black}{radioactive} source is not a point source. A typical source \textcolor{black}{assembly} we
envision is shown in
figure~\ref{source_weight_QC}. The source is enclosed in a 6~mm by 6~mm
cylindrical stainless steel shell, covered by \textcolor{black}{bullet-shaped} highly reflective
Polytetrafluoroethylene (PTFE), and attached to a stainless steel wire with 1~mm
diameter.
A PTFE connector 160~mm above the source allows easy exchange of
the source when desired. To maintain the tension in the wire, a weight of about 100~g
covered with PTFE is also attached to the wire below the source, with
a separation of about 160~mm.

Despite the PTFE materials, optical
photons can still be absorbed by these surfaces, leading to a small
bias in visible energy.
To study this bias,
a 90\% reflectivity is assumed for the PTFE surfaces
\textcolor{black}{based on the worst case in the laboratory measurement~\cite{private_comm1}}.
Only events with energy fully absorbed in the
LS region are chosen from the simulation to decouple this effect from
the energy loss in dead material (\textcolor{black}{see section~\ref{sec:eloss}}).

\begin{figure}
    \centering
    \includegraphics[width=3in]{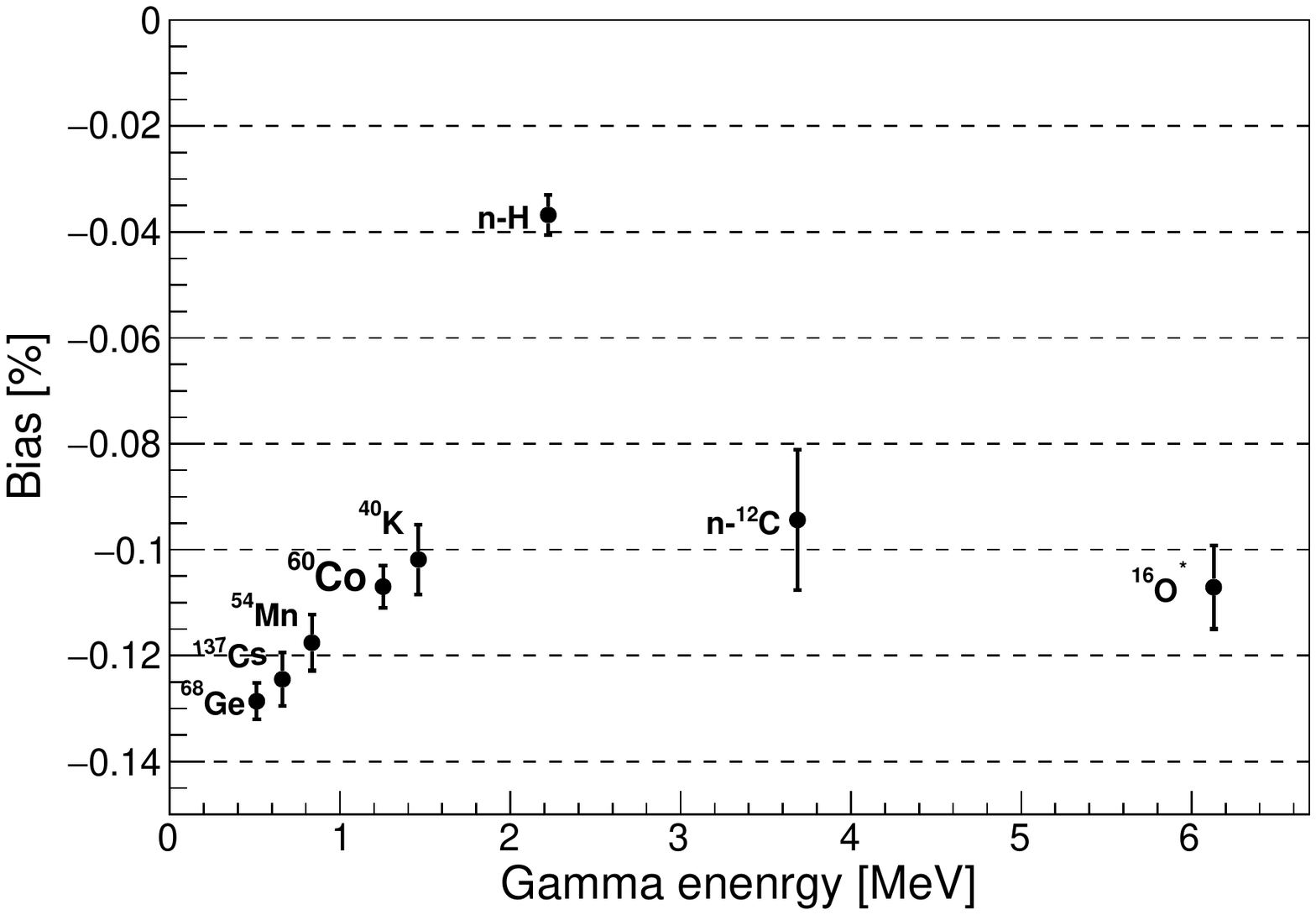}
    \caption{
    	\label{shadowing}
		Bias $(E_{\rm vis}^{\rm shadow} - E_{\rm vis}^{\rm ideal})/E_{\rm vis}^{\rm ideal}$ due to optical shadowing for individual gamma sources from the source assemblies as in figure~\ref{source_weight_QC}. \textcolor{black}{The lower probability of neutron capture on carbon than that on hydrogen leads to a large statistical uncertainty.}}
\end{figure}

The resulting biases, in
comparison to the bare sources in \textcolor{black}{figure~\ref{non_linearity_fit}},
are shown in figure~\ref{shadowing}.
The effect is less than 0.15\% \textcolor{black}{for} all sources.
\textcolor{black}{
The bias first reduces towards higher energy up to $^{40}$K (1.5 MeV), 
as these gammas deposit energy further away from the source enclosure.
The n-H point has the least bias due to additional displacement of
the neutron before its capture.
For the two highest energy gamma points, also produced by the neutron sources,
the photons are on average produced closer to the weight/connector and are being shadowed.}

\subsubsection{Energy loss effect}
\label{sec:eloss}
\begin{figure}
	\centering
    \includegraphics[width=3in]{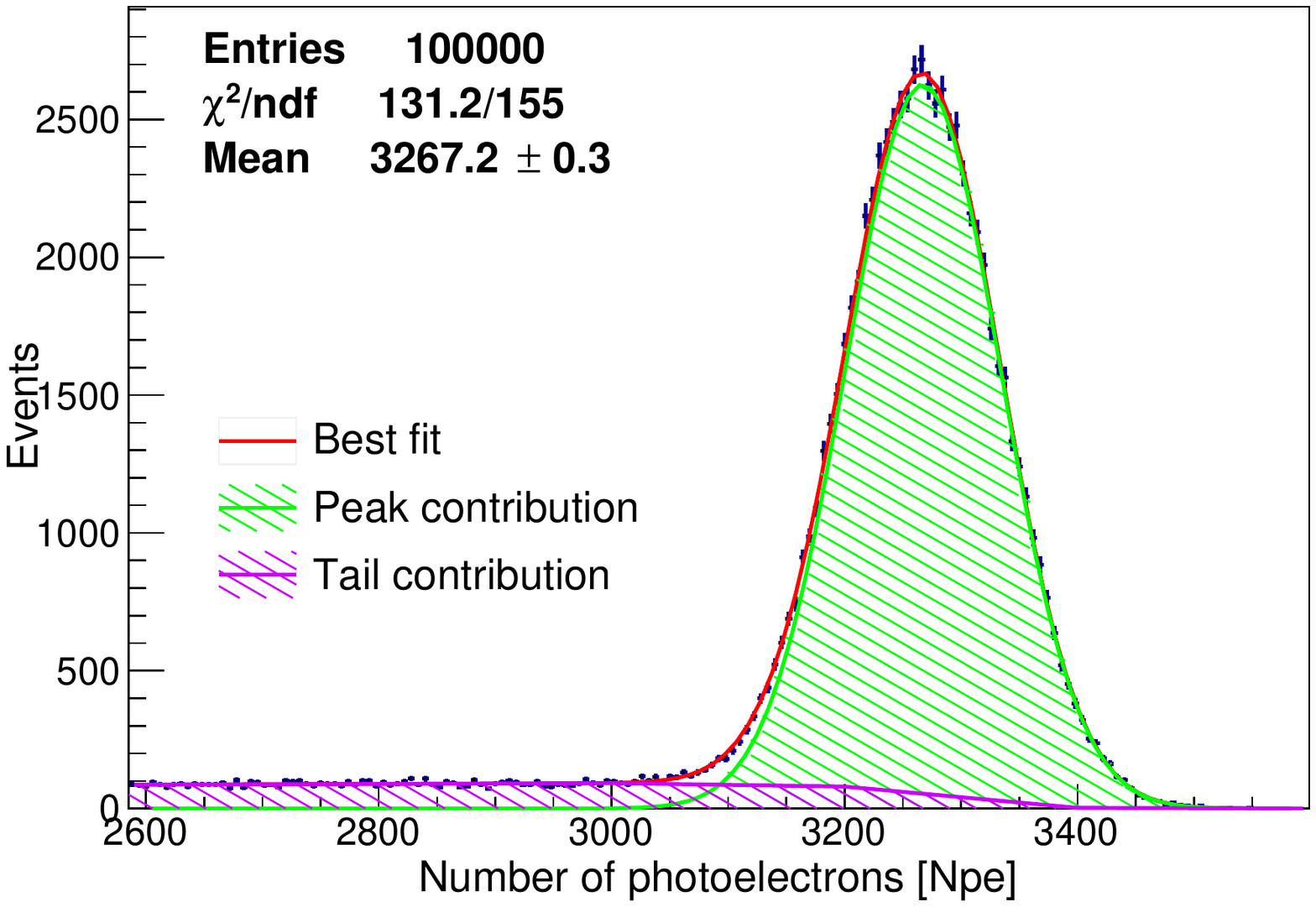}
    \caption{
    \label{Co60}
		The PE distribution of $^{60}$Co source at the
        center of JUNO. Contributions from the fully absorbed events
        and energy loss tail are indicated in the figure.}
\end{figure}

\begin{figure}
  \centering
  \includegraphics[width=3in]{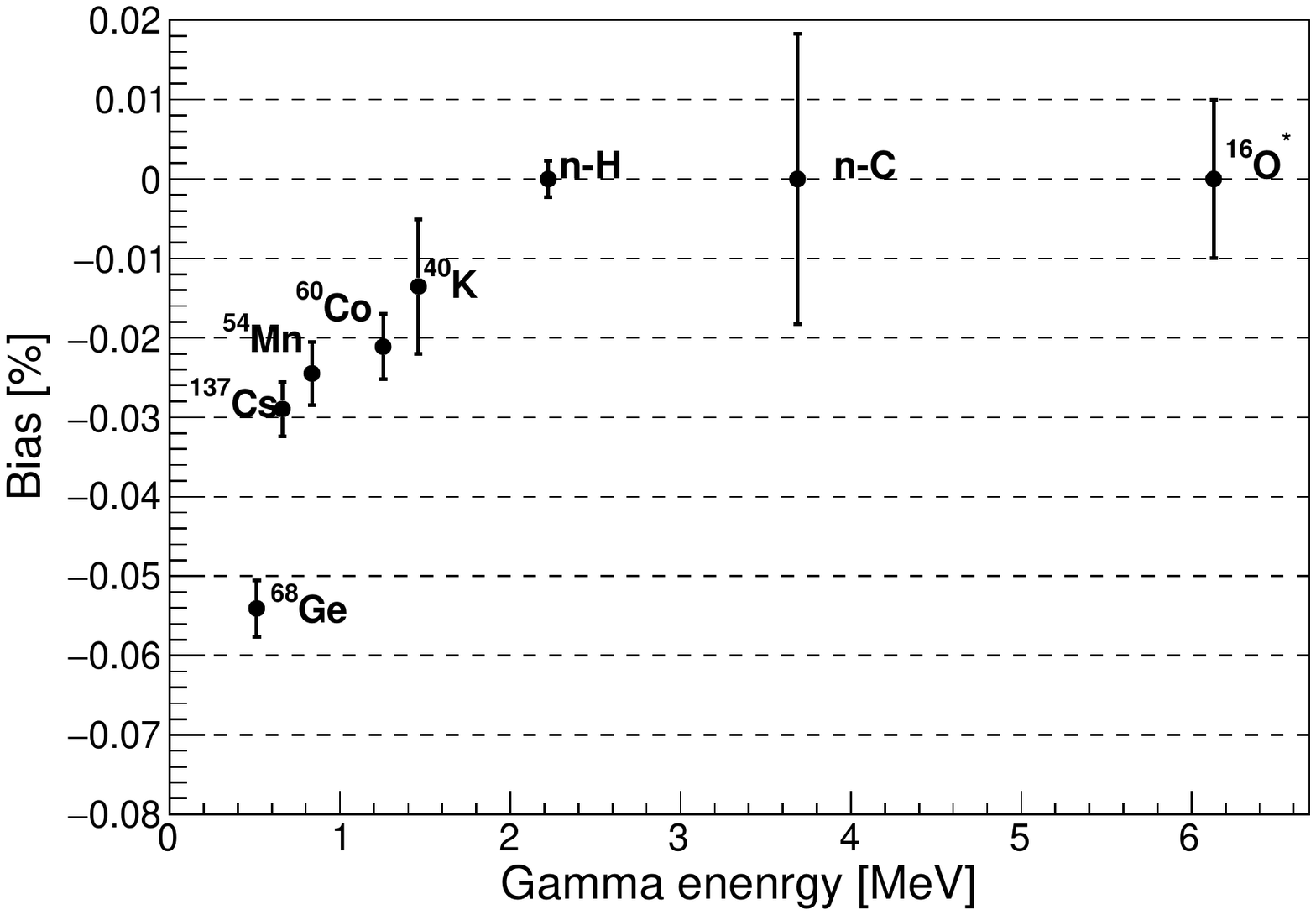}  
  \caption{
  	\label{compton}
	Residual fit bias $(E_{\rm vis}^{\rm loss} - E_{\rm vis}^{\rm ideal})/E_{\rm vis}^{\rm ideal}$  due to the energy loss effect for the gamma sources. 
    }
\end{figure}

Some gamma energy can also \textcolor{black}{be deposited} in the non-scintillating material,
e.g. the enclosure of the source, leading to a leakage tail in the
detected PE distribution. The bias to the peak is referred to as the
energy loss effect. As an example, the measured PE distribution
for a realistic $^{60}$Co source is shown in figure~\ref{Co60}, in
which contributions from the fully \textcolor{black}{absorbed peak in the LS} and the leakage tail
are separately plotted.
\textcolor{black}{A fitting function combining a single-value peak and an exponential tail,
convolved with a resolution function~\cite{fitmethod}, is adopted here}
to fit the simulated gamma
spectra. The fractional difference of the best fit peak to those from
the fully absorbed events are shown in figure~\ref{compton}, where
residual biases are less than \textcolor{black}{0.06\%} for all sources.

\subsubsection{PMT dark rate}
\label{sec:dark_rate}
Assuming an average dark rate of 30~kHz per LPMT channel (based on early quality assurance tests) and a 350~ns analysis time window, 
the dark rate pileup in any given physical event
is estimated to be \textcolor{black}{$\Delta Q_{\rm DR}$ = 350~ns$ \times $17600~LPMT$ \times $30~kHz/LPMT = 185~(PE) summed over LPMTs}.
Such an offset in energy can be precisely calibrated \textcolor{black}{and subtracted} using random triggers.

\subsubsection{\textcolor{black}{6.13 MeV gamma uncertainty}}
\label{sec:uncer_highE}$^{241}$Am-$^{13}$C source can produce a 6.13 MeV gamma,
but mixed with \textcolor{black}{neutron-proton recoil} energy. According to the
simulation, the positive bias caused by the neutron-proton recoil is at a
0.4\% level. Although subtractable, this leads to an uncertainty
at the high energy end.

\subsubsection{Instrumental non-linearity}
\textcolor{black}{As discussed in section~{\ref{sec:elec_nonlin}},
the uncertainty of the instrumental non-linearity can be controlled below 0.3\%, 
conservatively taken as a fully correlated uncertainty in energy.}

\subsubsection{\textcolor{black}{Residual bias after non-uniformity correction}}
\label{sec:position_dep_bias}
\textcolor{black}{Since IBDs are detected throughout the detector,
position-dependent energy scale variation has to be corrected for.
The correction will be discussed in section~{\ref{sec:resolution}}
as a major ingredient of energy resolution optimization.
The residual bias is controlled to 0.3\% level,
which is conservatively taken as a fully correlated uncertainty in energy.}


\subsubsection{Combined systematic uncertainty}
\label{sec:sys_uncer}
The effects discussed above are summarized in
table~\ref{summary_uncertainty}. We assume that all the \textcolor{black}{biases} can
be corrected, but with conservatively 100\% uncertainty. We further separate the
uncertainties into either correlated \textcolor{black}{at} different energy, or single point,
\textcolor{black}{depending on how they would move individual energy points up and down.}
For example, the uncertainties due to the shadowing effects
are correlated among different sources. On the other hand, 
the bias due to 6.13 MeV gamma is independent from the others (single point).

  To evaluate the combined effects, mock calibration data are produced by randomly
  biasing the naked source values (figure~\ref{non_linearity_fit}) and $^{12}$B visible energy spectra (figure~\ref{B12})
  according to the 1$\sigma$ uncertainties in
  table~\ref{summary_uncertainty}, either in a correlated or
  single point fashion. 
  For each set of the data,
  a fit as in section~{\ref{sec:calib_proc}} is performed, yielding an
  electron non-linearity curve. This is repeated many times.
  The individual contribution is obtained by ``turning-on'' only one effect in table~\ref{summary_uncertainty} at a time
  \textcolor{black}{and by repeating} the procedure.
  The dominating contribution comes from instrumental non-linearity and position-dependent effect,
  each contributing to 0.3\% in the overall uncertainty band.
  The 1$\sigma$ distribution of these fitted models are overlaid in
  figures~\ref{non_e_linearity_fit}
  and~\ref{fig:positron_non-linearity}. The uncertainty band is
  $\sim$0.7\% across all energy,
  \textcolor{black}{which is better than the 1\% requirement.
  Note that such precision has been experimentally corroborated by
  Daya Bay~\cite{dayabay-calib-paper} and Double Chooz~\cite{Double-Chooz-nature-physics}.}
\begin{table}[H]
    \footnotesize
	\centering
    \begin{tabular*}{150mm}{@{\extracolsep{\fill}}cccc}
          \hline 
		  source & bias & uncertainty & nature\\
          \hline
          Shadowing effect & $-$0.1\% - $-$0.2\% & 0.1\% - 0.2\% & correlated \\
          Energy loss effect & $-$0.05\% - $-$0.1\% & 0.05\% - 0.1\%  & correlated \\
          6.13 MeV gamma uncertainty & $+$0.4\% & 0.4\% & single point \\
          Instrumental non-linearity & n/a & \textcolor{black}{0.3\%} & correlated\\
          Position-dependent effect & \textless0.3\% & 0.3\% & correlated\\
          \hline  
    \end{tabular*}
    \caption{
        \label{summary_uncertainty}
		A summary of the systematic uncertainties in energy scale.
    The uncertainties are set equal to the absolute values of
    the bias for conservativeness.
    The last column indicates \textcolor{black}{whether} this uncertainty is
    correlated among different sources or energy,
    or should only be applied to a single point.
    }
\end{table}

\section{Optimizing the energy resolution}
\label{sec:resolution}
\begin{figure}
        \centering
        \includegraphics[width=3in]{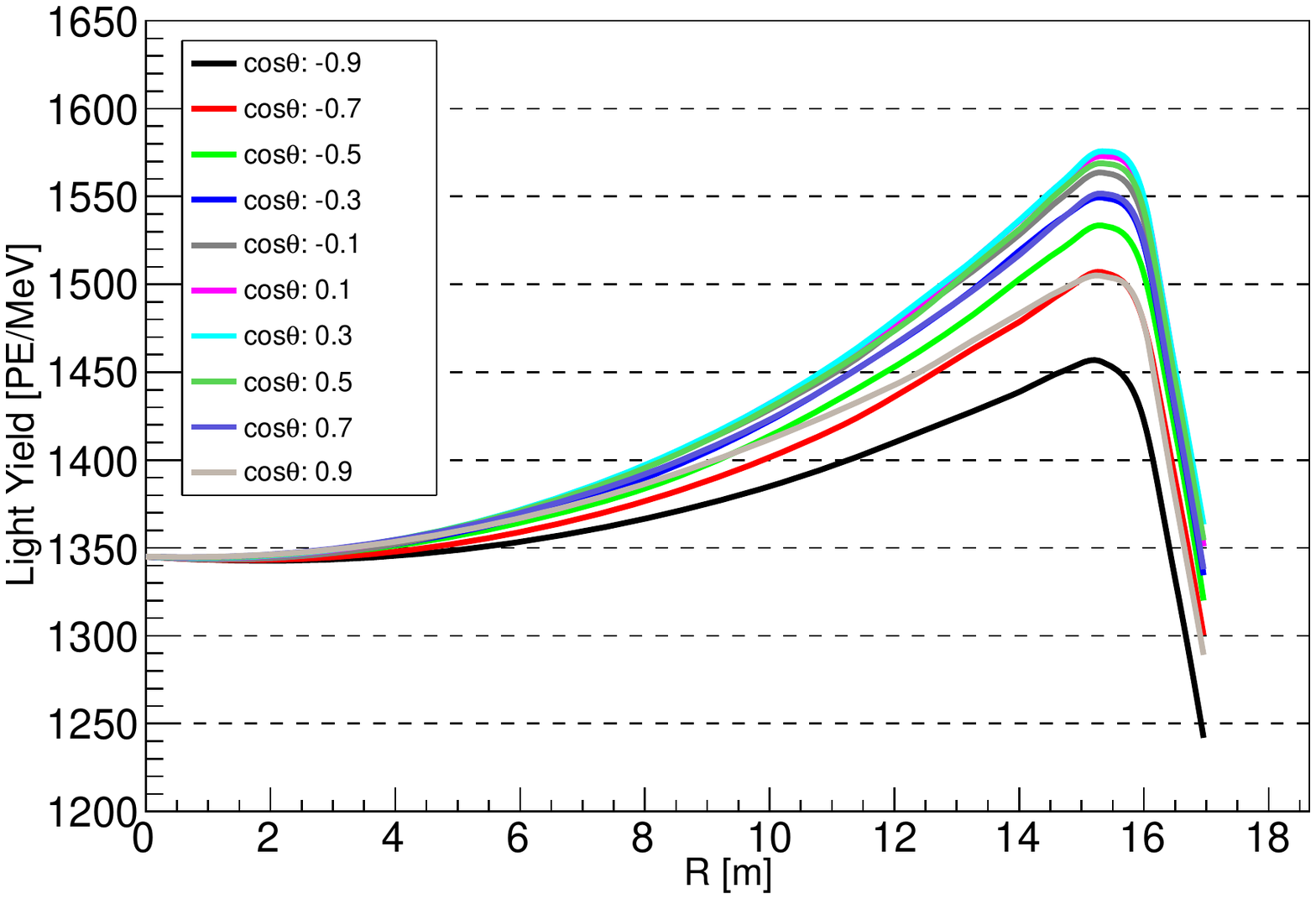}
        \caption{	
        	\label{PE_vs_R}
			Mean number of PEs \textcolor{black}{per MeV} for 2.22 MeV gammas
        as a function of the radius,
        along a few representative polar angles.}
\end{figure}

\textcolor{black}{As outlined in the introduction, another} key aspect of the calibration is to \textcolor{black}{optimize} the energy
resolution for the IBD positron signals. In general, the fractional
energy resolution for a visible energy $E_{\rm vis}$ in MeV can
be written as an approximate formula
\begin{eqnarray}
  \label{eq_resolution}
  \textcolor{black}{
  \frac{\sigma_{E_{\rm vis}^{\rm prompt}}}{E_{\rm vis}^{\rm prompt}}=\sqrt{\left(\frac{a}{\sqrt{E_{\rm vis}^{\rm prompt}}}\right)^{2}+b^{2}+\left(\frac{c}{E_{\rm vis}^{\rm prompt}}\right)^{2}}\,.
  }
\end{eqnarray}
The $a$ term is the statistical term, which is mainly driven by the
Poisson statistics of true number of PEs associated with $E_{\rm vis}$.
\textcolor{black}{To set the scale, for a light yield Y$_{0}$ of 1345~PE/MeV, $a$ is about 2.7\%.}
The $b$ term is a constant independent of
energy, dominated by the position
non-uniformity. In figure~\ref{PE_vs_R}, the average number of PEs for
2.22 MeV gammas vs. radius is shown for a few representative
polar $\theta$ angles in spherical \textcolor{black}{coordinates}. Starting from the detector
center, the gradual increase is a combined effect of variations
in \textcolor{black}{active photon coverage} and the attenuation of optical photons. The sharp decrease close to 15.5~m
is due to the mismatch of the indices of refraction between the
acrylic and water -
the closer the event to the edge,
the more likely is the occurrence of total reflection and consequently,
the loss of photons by absorption.
\textcolor{black}{Within the same radius, there is an additional dispersion
in $\theta$ due to effects such as PMT coverage and photon shadowing on opaque materials, etc.}
The parameter $b$ estimated from figure~\ref{PE_vs_R}
is of the order 10\%, but can be largely \textcolor{black}{suppressed}
based on position-dependent calibration. The $c$ term represents the contribution of a background noise, i.e. the
dark noise from the PMTs, which is always mixed with $E_{\rm vis}$ in
the measurement. The charge bias induced by dark rate ($\Delta Q_{\rm DR}=185$~PE, see section~\ref{sec:dark_rate})
has a Poisson noise of 13.6 PE, so $c$ is estimated to be 1.0\%.
\textcolor{black}{The impact of the energy resolution
Eq.~\eqref{eq_resolution} to the MO determination has been
studied in a previous publication~\cite{yellow-book} with the following procedure.
Mock neutrino energy spectra were generated assuming different
values of $a$, $b$ and $c$ in Eq.~\eqref{eq_resolution},
based on which MO fits were performed to calculate the medium MO sensitivity.
It was found, numerically, that the JUNO baseline requirement to determine the MO to
3~--~4 $\sigma$ significance could be translated into
a convenient requirement on an effective resolution $\tilde{a}$ as:}
\begin{eqnarray}
\label{eq_overall_res}
\tilde{a} \equiv \sqrt{(a)^{2}+(1.6 \times b)^{2}+\left(\frac{c}{1.6}\right)^{2}} \leqslant 3 \%\,.
\end{eqnarray}
Conceptually, the facts that in Eq.~\eqref{eq_resolution} the second term is not improving
with the visible energy, and that the third term 
declines quickly with the energy, are reflected in factors 1.6 and
1/1.6, respectively. Note also that $\tilde{a}$ is only an effective
parameter and is not the detector energy resolution at 1 MeV.

\textcolor{black}{From the perspective of calibration,
reducing the $b$ term is the key to optimize the energy resolution,
particularly given the large apparent non-uniformity depicted in figure~\ref{PE_vs_R}.}
The non-uniformity can be studied by either deploying radioactive sources to fixed locations, or by using uniformly distributed
background events, e.g. spallation neutrons
($\sim$1.8~evt/s~\cite{yellow-book}). Although we focus on the first
approach in this paper, it should be emphasized that the second
approach \textcolor{black}{will} offer a powerful {\it in situ} cross check.

\begin{figure}
    \centering
    \includegraphics[width=3in]{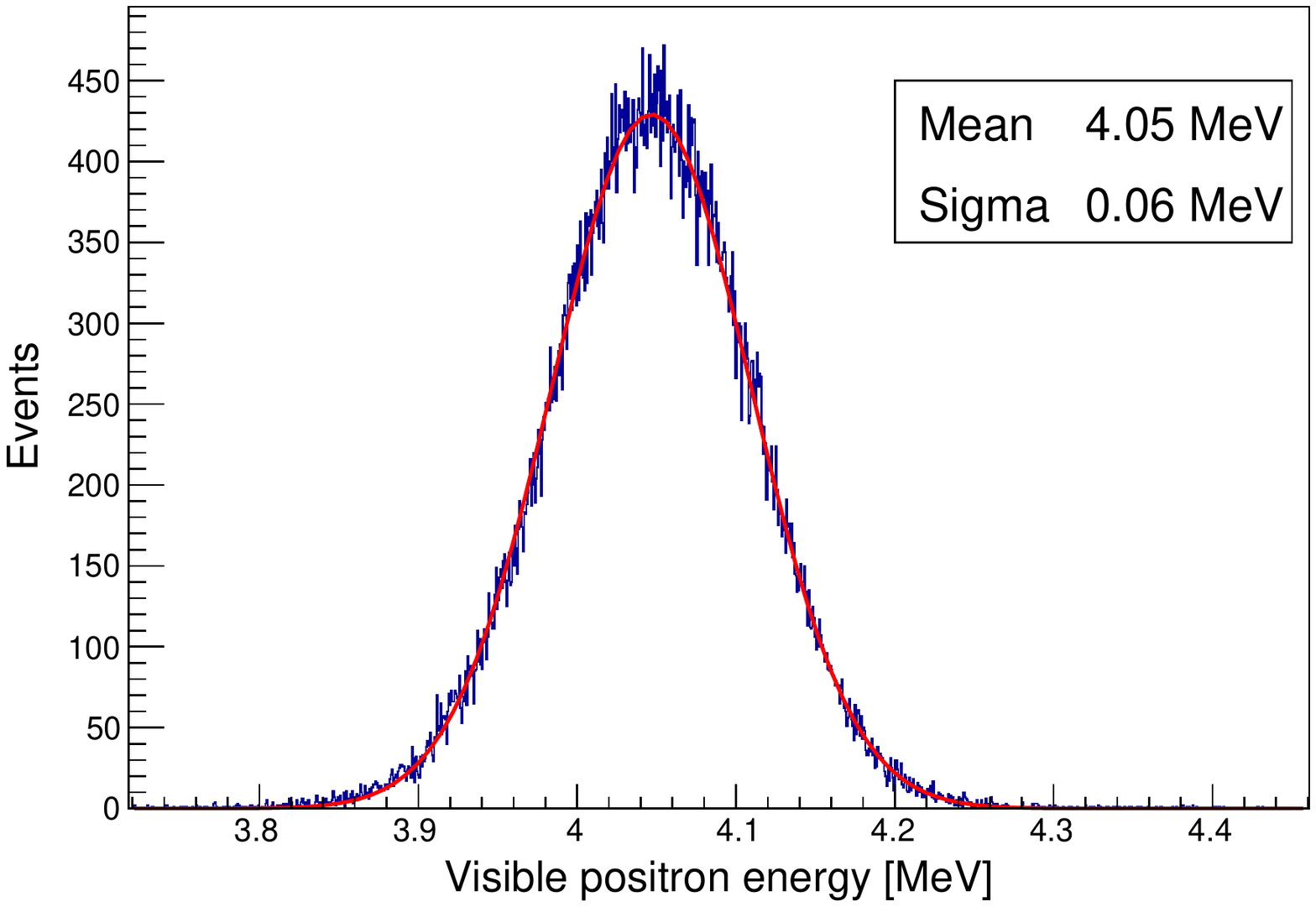}
    \caption{
    	\label{gauss_fit}
	    A typical Gaussian fit to obtain the energy resolution $\sigma$/mean for 3 MeV positrons, with $E^{\rm prompt}_{\rm vis}$ reconstructed by Eq.~\eqref{eq_E_rec}.}
\end{figure}

The non-uniformity is characterized by $g(r,\theta,\phi)$, defined as the light
yield in a given position relative to that at the center. The key question is how to 
calibrate $g$ sufficiently well under realistic situations. 
In this study, we start by assuming that we have perfect positron sources deployable to any given location in the detector. 
We then gradually go to more realistic calibration with gammas 
at finite points.
The performance of the calibration is assessed by reconstructing the visible energy for uniformly distributed IBDs 
at each MeV from 0 to 8 MeV as:
\begin{eqnarray}
\label{eq_E_rec}
\textcolor{black}{E_{\rm vis}^{\rm prompt}(r,\theta,\phi)} = ({\rm PE_{\rm tot}}-\Delta Q_{\rm DR})/Y_0/g(r,\theta,\phi)\,,
\end{eqnarray}
\textcolor{black}{where $\rm PE_{\rm tot}$ is the total number of PEs for the LPMT and SPMT, and the dark rate pileup $\Delta Q_{\rm DR}$ is included as an independent Poisson-fluctuated offset. }
After obtaining the resolution $\sigma$/E for each energy via Gaussian fit as in figure~\ref{gauss_fit},
($a$, $b$, $c$) can be extracted and compared to Eq.~\eqref{eq_overall_res}.

\subsection{Central IBDs}
\label{central_IBD}
\textcolor{black}{To start, IBDs are produced from the geometrical
center of the detector,
so that the position non-uniformity is completely absent, i.e. $g=1$.
However, the average number of detected PEs is lower than the full volume (figure~\ref{PE_vs_R}).}
Interestingly, a fit of the energy resolution of these positrons using Eq.~\eqref{eq_resolution}
yields $a=2.62\%$ and a non-vanishing $b=0.73\%$. The parameter $c =
1.38\%$ is also sizably larger than the naive dark rate estimate
(1.0\%).  The origins of these non-Poisson fluctuations are traced in
the simulation by switching off individual processes. The residual $b$
is dominated by non-Poisson distribution of Cherenkov photons
due to track length fluctuations,
leading to additional ``constant'' noise term in the detected PEs.
The additional contribution in $c$ is due to the fluctuations
in secondary electrons produced by the annihilation gammas,
folded with their corresponding non-linearity,
leading to a smearing effect independent of the positron energy.

\subsection{Ideal non-uniformity correction}
In the ideal case, uniformly distributed IBDs are the calibration sources themselves.
The detector is divided into 20,000 equal-volume \textcolor{black}{``voxels''}, and
$g(r,\theta,\phi)$ is computed in each voxel for every energy,
allowing a potential energy dependence \textcolor{black}{to take into account effects such as energy leakage at the edges.}

\begin{figure}
    \centering
    \includegraphics[width=3in]{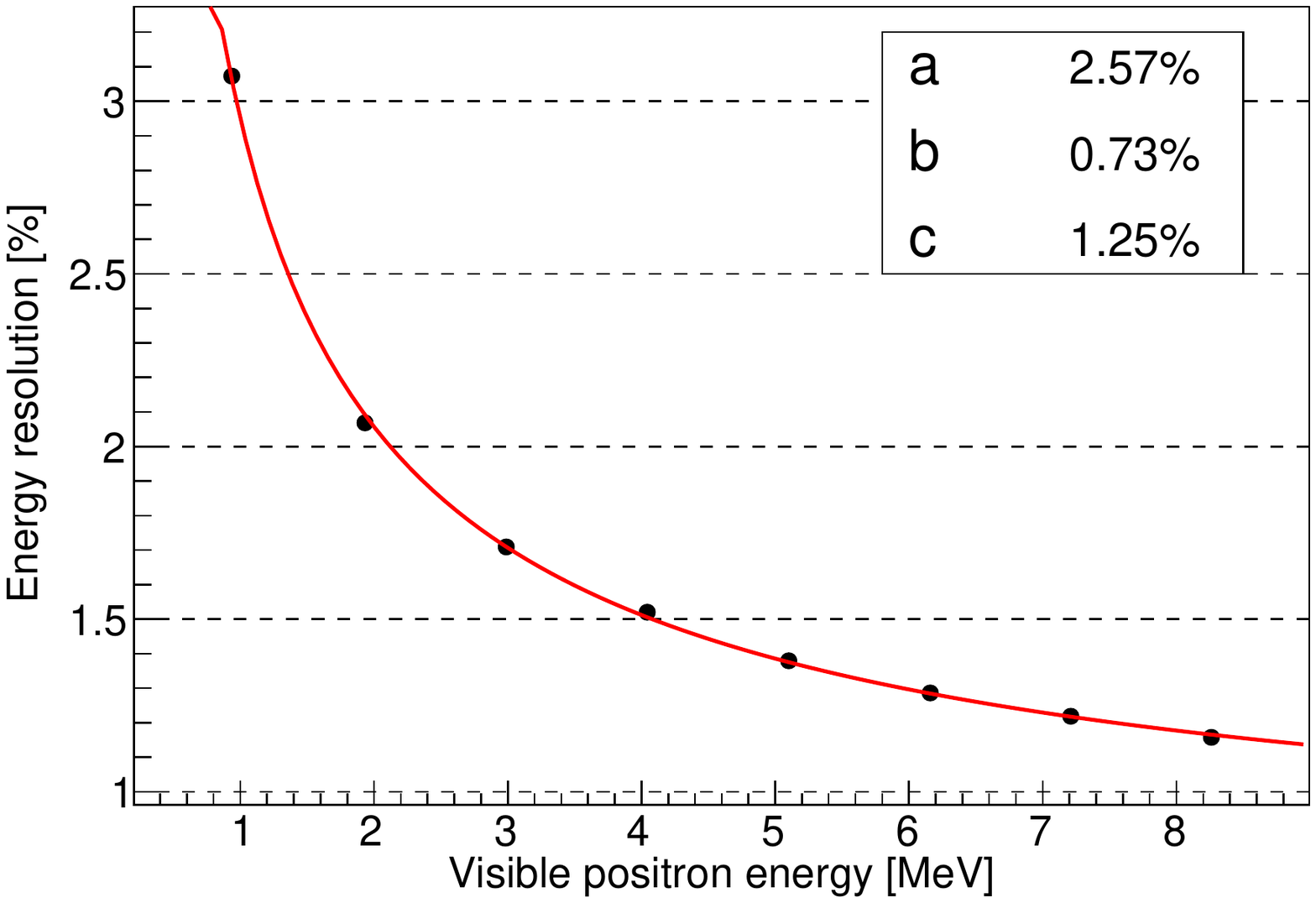}
    \caption{
    	\label{cls_position_smearing}
		Positron energy resolution vs. $E_{\rm vis}^{\rm prompt}$ after
          the ideal non-uniformity correction (uncertainties are smaller than the markers). 
          The red line shows the fit using Eq.~\eqref{eq_resolution}. }
\end{figure}

The fit of the energy resolution is shown in
figure~\ref{cls_position_smearing}, yielding
$a$=2.57\%, $b$=0.73\%, $c$=1.25\%, and $\tilde{a}=2.93\%$.
In comparison to those of the center IBDs,
the reductions in $a$ and $c$ terms are due to the increase of
full-volume \textcolor{black}{detected PEs} (figure~\ref{PE_vs_R}). This scenario will
serve as the most ideal reference for the calibration.

\subsection{Utilizing azimuthal symmetry}
\begin{figure}
    \centering
    \subfigure[]{
        \includegraphics[width=2.8in]{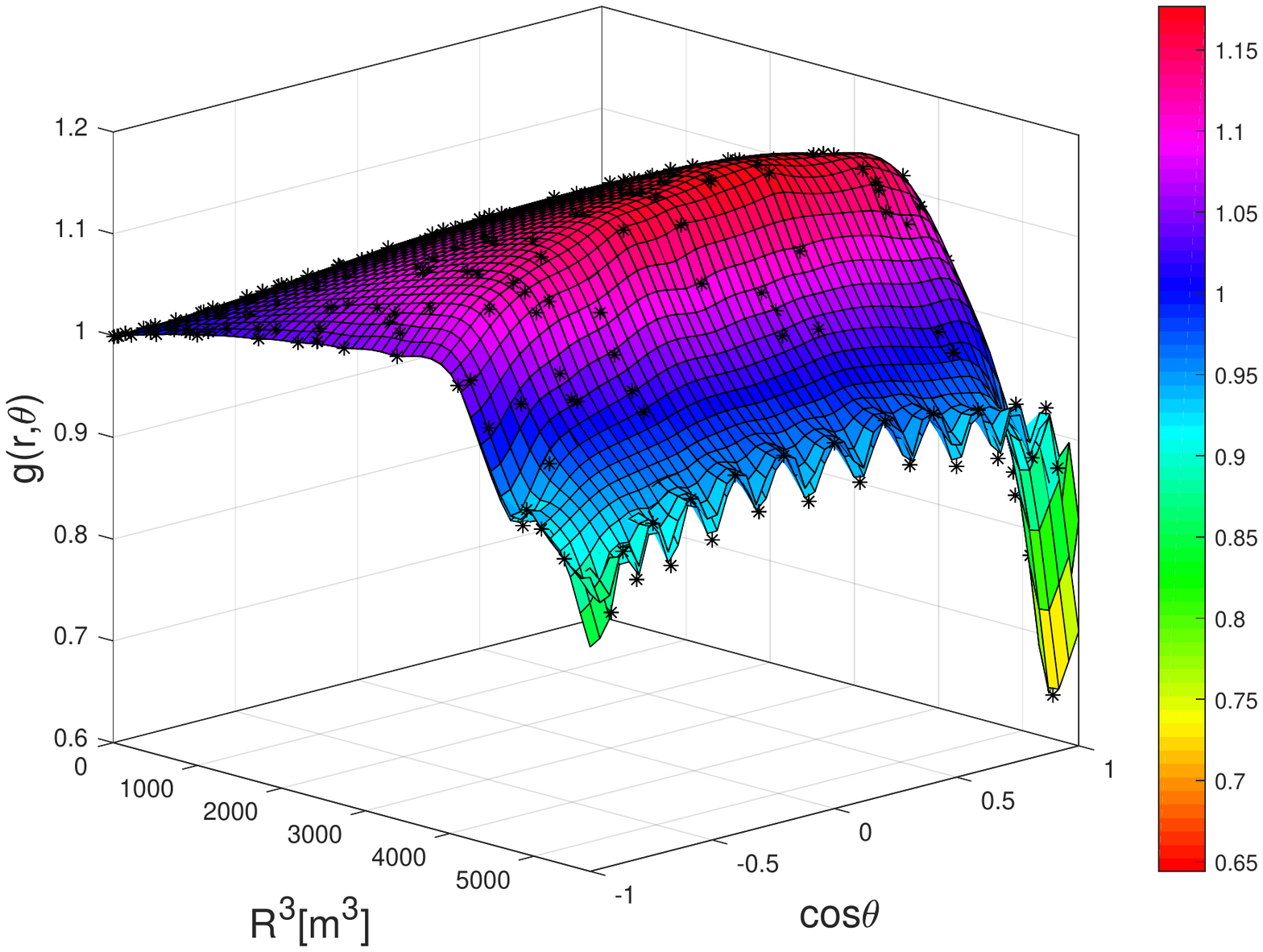}
        \label{spline_function_3D}
    }
    \subfigure[]{
        \includegraphics[width=2.8in]{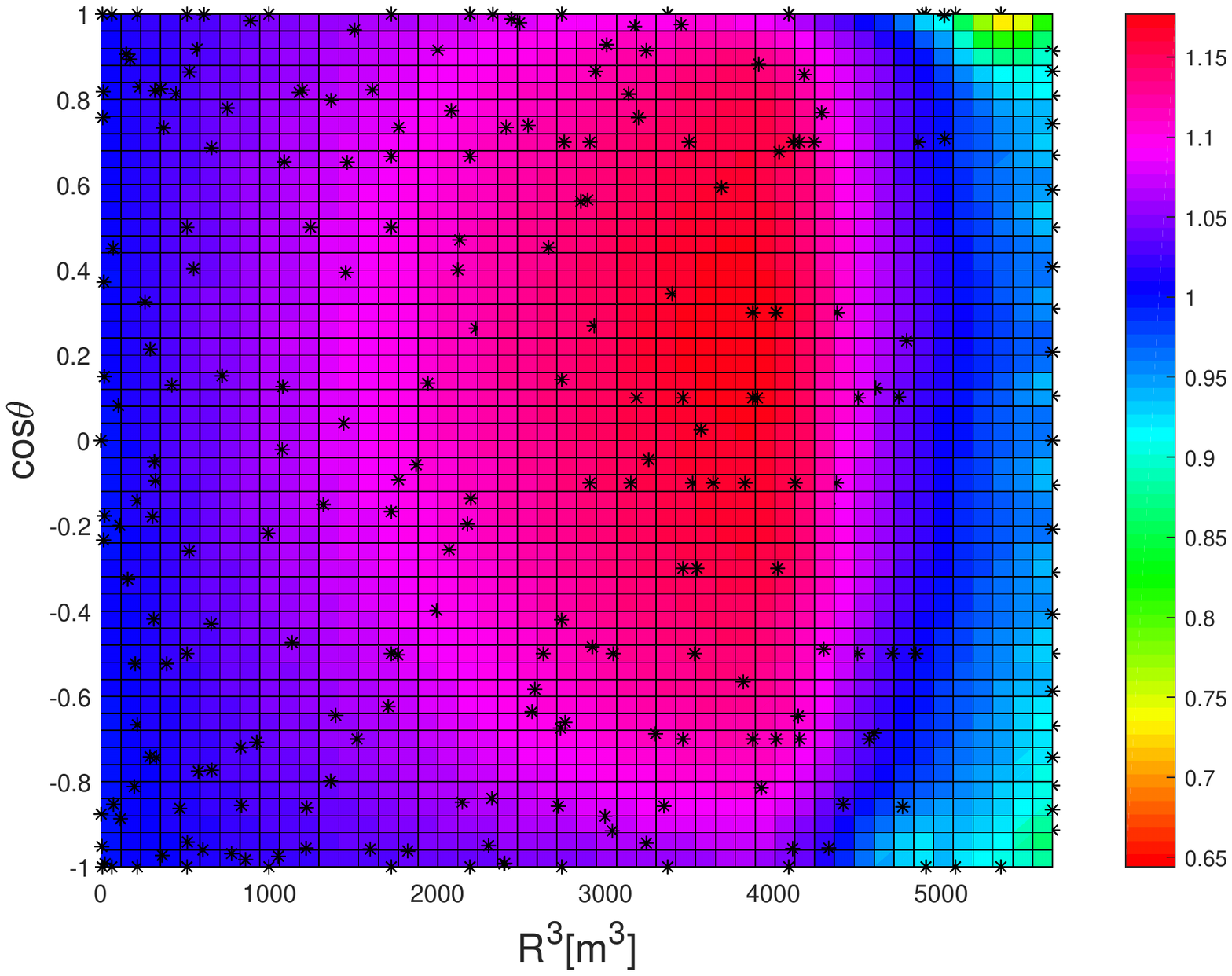}
        \label{spline_function_2D}
    }
    \caption{
        The non-uniformity $g(r,\theta)$ in $\phi=0$
          plane as a function of $\cos\theta$ and $R^{3}$, 
          (a) 3D surface, (b) 2D projection with value of $g$ indicated by the color band. 
          The black points are the optimized deployment locations of calibration sources
          as explained in section~\ref{finite-calib-p}.
        }
    \label{spline_function}
\end{figure}

The azimuthal symmetry in JUNO detector allows the first
simplification that only $g$ in a vertical plane of the detector
is needed. For illustration, $g(r,\theta)$ surface in the $\phi=0$
plane for positrons with zero kinetic energy is shown in
figure~\ref{spline_function}. The two-dimensional $g(r,\theta)$ function
obtained at each positron energy is then applied to the corresponding
positron events.
The resulting $\tilde{a}$ is 2.96\%.

\subsection{Single gamma source}
\textcolor{black}{In this step, the hypothetical positron source
is replaced by a realistic AmC neutron source (2.22 MeV capture gammas)}, so that both the 
particle type and energy dependence in $g(r,\theta)$ are neglected.  
The parameter $\tilde{a}$ is found to be 2.98\%. Note that this choice 
would also allow direct comparisons to the spallation neutron capture signals in the
experiment.

\subsection{Finite calibration points}
\label{finite-calib-p}
\begin{figure}
  \centering
  \includegraphics[width=2.8in]{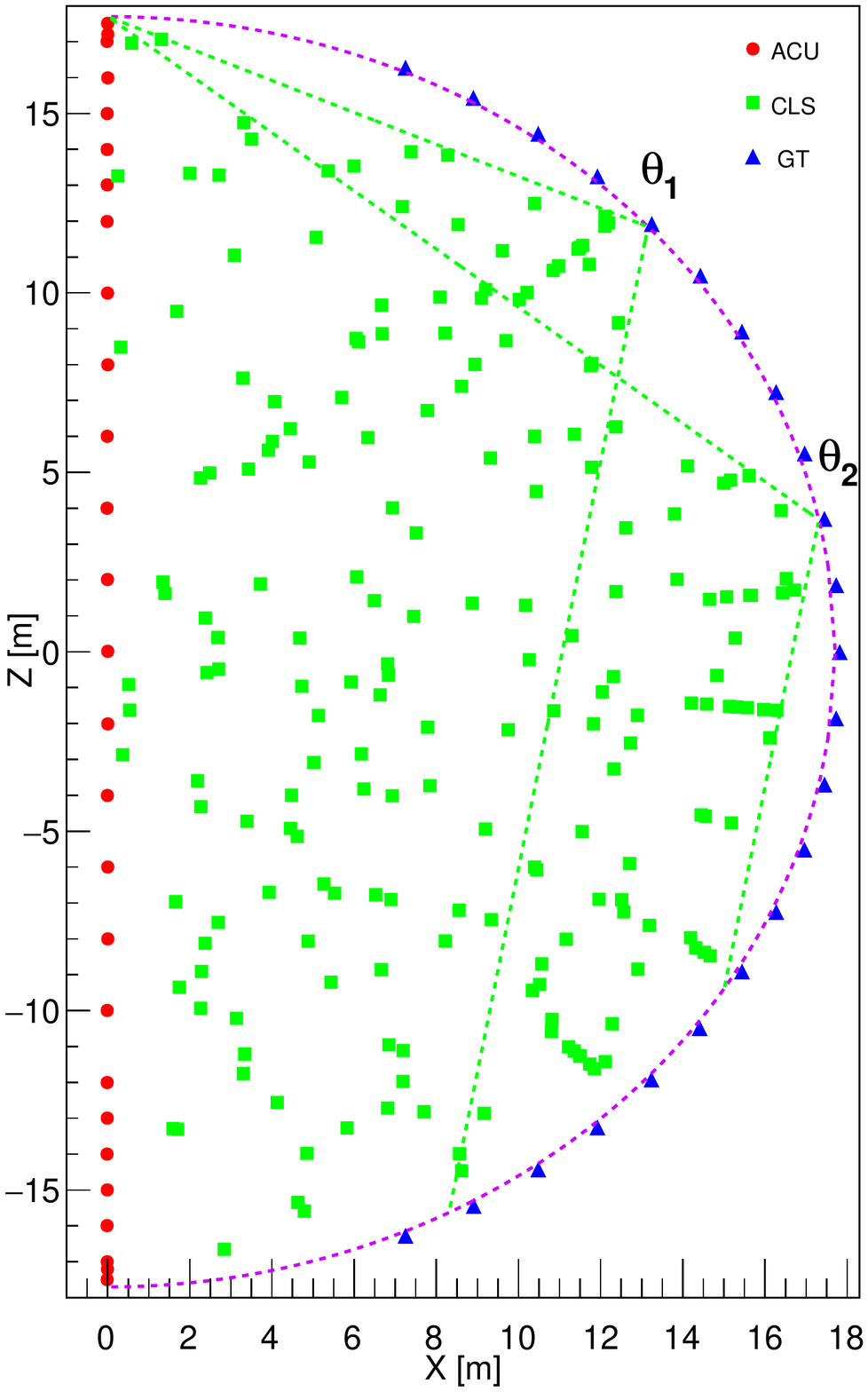}
  \caption{
    \label{basic_coverage}\textcolor{black}{
	The optimized 250 calibration points in a half vertical plane of the JUNO CD, 
	based on the procedure in the text. 
	The purple dashed line represents the acrylic boundary. 
	The ACU, CLS and GT are individual source deployment systems discussed later in section~\ref{sec:CalibSys}. 
	The green dashed lines are the assumed boundaries of the CLS.}}
\end{figure}

The next critical \textcolor{black}{step} is to select a list of ``must-do''
calibration points while still maintaining a good approximation for $g(r,\theta)$.
Several pragmatic considerations are made:
\begin{enumerate}
\item As observed in figure~\ref{spline_function}, there is no apparent
  symmetry in $g(r,\theta)$ that allows a vast simplification to the
  choice of points. However, the variation of $g$ appears faster at
  the edge, which mandates more sampling points there.

\item To constrain the surface of $g(r,\theta)$, the boundaries of the
  vertical half-plane are critical, i.e., the central vertical axis of the
  detector and the circle at the LS-acrylic boundary.  The
  central axis can be calibrated with good granularity,
  one point every 2 meters for radius less than 12 m, one point every
  meter from 12 m to 17 m, and two points at 17.2 m and 17.5 m (27
  points in total).  For the acrylic boundary, the sphere is made up
  in 23 layers in vertical direction, each supported by stainless steel fixtures with non-trivial
  impact to the optics. It would therefore be sensible
  \textcolor{black}{to do calibration in the vicinity of these locations.}

\item \textcolor{black}{For the region in between the central axis and
    acrylic boundary, 200 points would be reasonable,
    e.g. selecting 20 points along each of the radial line in figure~\ref{PE_vs_R}.
    In this study, however, a semi-random approach is adopted to choose these
    200 points in order to yield the best calibration performance.
    Since the response varies fast at large radius,
    300 regular points are first selected along the 10 radial lines
    between 15 and 17 m, then another 1000 random points are chosen
    in the vertical plane. This 1300-point grid is used as the 
    basic template from which the optimal 200 points are decided (see later).}

\item Realistically, not all points in $(r,\theta)$ are accessible. If
  we consider a cable loop system with an anchor on
  the inner surface of the CD~\cite{JUNOCDR}, some points too close to
  the vertical geometrical limit cannot be reached due to the loss of
  cable tension (ref.~\cite{zhangyuanyuan-new-paper}). To allow a
  good coverage we consider two cable loops
  (figures.~\ref{basic_coverage}~and~\ref{overview_calibration_system}) in the
  CD. Note that physically the two loops can be at different vertical
  planes, and azimuthal symmetry can be applied to combine them into a
  single $(r,\theta)$ half-plane.
\end{enumerate}
\textcolor{black}{The choice of the calibration positions is correlated
with the choice of the two anchor locations ($\theta_1$, $\theta_2$)
of the cable loops. The following automatic optimization procedure is performed:}
\begin{enumerate}
    \item \textcolor{black}{The 27+23 points are fixed along the vertical central axis and the LS-acrylic boundary.}
    
    \item \textcolor{black}{All possible ($\theta_1$, $\theta_2$)
    combinations are considered}, each with its own accessible region in $(r,\theta)$
    (figure~\ref{basic_coverage}).
    
    \item \textcolor{black}{For each ($\theta_1$, $\theta_2$), 200
    points are randomly chosen from the 1300-grid template.}
    
    \item For each set of 200 (random) plus 50 (fixed) points, simulations with 2.22~MeV
    gammas are performed. A smooth
    surface of $g(r,\theta)$ is constructed using a two-dimensional
    spline function.
    
    \item $g(r,\theta)$ thus obtained is applied to the uniform
    positron events, on which $\tilde{a}$ is extracted as a
    figure-of-merit. Steps 1 through 5 are repeated until a minimal
    $\tilde{a}$ is found.
    
\end{enumerate}
The optimal choice of the calibration points are
illustrated in figure~\ref{basic_coverage}, in which the best anchor
locations are $\theta_1=\ang{48}$ and $\theta_2=\ang{78}$, and the resulting $\tilde{a}$ is 2.98\%.
\textcolor{black}{On the other hand, for this $(\theta_1,\theta_2)$, out of 100,000 random choices, 40,000 choices yield $\tilde{a}$ less than 3.0\%.}

\subsection{Vertex smearing}
\textcolor{black}{So far, true production vertex of the IBDs have been used to look up $g(r,\theta)$.}
In reality, the event-by-event correction has to rely on the reconstructed interaction vertex,
which will in turn \textcolor{black}{affect} the quality of the correction.  An
algorithm based on time and charge information of LPMTs \textcolor{black}{has} been
developed in Ref.~\cite{vertex-rec}, where a resolution of
\textcolor{black}{8~cm/$\sqrt{E(\rm{MeV})}$} (average distance between the true and
reconstructed vertex) was achieved. To study the impact to the overall energy
resolution, three vertex resolution assumptions have been made here, 
8, 10, and 15~cm/$\sqrt{E(\rm{MeV})}$. 
For each simulated event, Gaussian position smearing is applied accordingly before applying
$g(r,\theta)$ in section~{\ref{finite-calib-p}}. The corresponding \textcolor{black}{result for}
$\tilde{a}$ is 3.01\%, 3.05\%, and 3.10\%, respectively.

Another related effect is \textcolor{black}{the uncertainty on} the calibration source
location, leading to a small smearing
on $g(r,\theta)$ values. The hardware requirement is a precision of
3~cm~\cite{JUNOCDR} - significantly better than the reconstructed
\textcolor{black}{spatial} resolution at 1~MeV. It is verified that this contribution can be
neglected.

\subsection{LPMT quantum efficiency variation}	
  For each LPMT, the quantum efficiency (QE) was
  measured individually at 430 nm during the quality assurance process, and 
  can be approximated by a Gaussian with a mean of $\sim$30\%, and a $\sigma$/mean of 
  7\%~\cite{man_in_prep}.
  \textcolor{black}{To take this effect into account, QEs of individual LPMTs in the simulation are drawn from the Gaussian distribution. The resulting $\tilde{a}$ is 3.02\%.}

\subsection{Realistic detector}
\label{sec:real_det}
So far we have discussed the simulation of an ideal JUNO detector with nominal
parameters in the simulation. A real detector can certainly differ from the
simulation, so it is important to study whether an {\it in situ}
calibration in a realistic detector can still achieve the required
resolution. Five conservative alterations of the CD are considered:
\begin{enumerate}
\item The JUNO LPMTs and readout electronics are designed to yield less
  than 1\% dead channels ($\sim$180) in six years. We assume a CD
  with 1\% LPMT failure in random positions. 

\item Same as above but forcing an additional asymmetry among the bad
  channels \textcolor{black}{(75:105, 
  an unlikely asymmetric failure with only 1\% p-value if individual PMTs are failing randomly)} 
  in the two semispheres separated by the vertical calibration plane, 
  breaking the assumed azimuthal symmetry.

\item The JUNO LS has been tested in one of the
    decommissioned Daya Bay detector for 1.5 years, so all optical
    parameters in JUNO LS are validated with \textcolor{black}{these} data. No temporal
    degradation of the light yield has been
    observed with \textcolor{black}{a maximum variation of $\pm$0.5\%} among
    measurements~\cite{YuZeyuan_LSpaper}. \textcolor{black}{To be conservative}, we
    assume that the light yield $Y_0$ is reduced by 1\%
    and 5\% and study their effects.

\item A 4\% reduction of the absorption length (\textcolor{black}{default} 77~m at
  430~nm~\cite{JUNOLSAbsL}) can also produce a 1\% reduction
  of $Y_0$. However, instead of a global reduction of light yield
  everywhere, the reduction of absorption length alters the uniformity
  of the detector. We assume such scenario in the simulation.
  Note that changes in LS transparency (both absorption and scattering) 
  will be monitored \textcolor{black}{to a precision of 1~m} with 
  a laser system \textcolor{black}{named} AURORA (figure~\ref{overview_calibration_system}).

\item The average single PE resolution from the bench measurement of LPMTs is 30\%~\cite{man_in_prep}, which will also affect the
    energy resolution, \textcolor{black}{especially if the integral
    of the waveform is used as the energy estimator. A 30\% single PE charge smearing is applied to 
    each of the simulated PE from the LPMTs.}
\end{enumerate}

In each case,
\textcolor{black}{$g(r,\theta)$ is calibrated with the 250 points in figure~\ref{basic_coverage}, 
and then gets applied} to uniformly distributed
positrons. The results are summarized in
table~\ref{summary_resolution}. For the two cases
  where $\tilde a$ exceed 3.1\% (5\% reduction of light yield, or the
  inclusion of 30\% single PE smearing), sizable increases
  of the $a$ term are observed (as expected), which obviously could not 
  be mitigated by the calibration. \textcolor{black}{For the 30\% single PE resolution, 
  our treatment is conservative since
	within the IBD energy range, 
	there are still 40\% of the LPMTs working under the single 
	photon regime, the counting of which could partially mitigate the charge smearing.
	A successful application of this approach is in
    ref.~\cite{Lux-single-photons}. A waveform deconvolution
    technique~\cite{huangyongbo-papper} may also offer a better
    estimator for the number of PEs.} 

\subsection{Bias in the energy scale}
  \label{sec:bias}
  A residual bias in positron energy scale can still exist after the non-uniformity correction. 
  With nominal JUNO detector and 250-point-based $g(r,\theta)$,  
  \textcolor{black}{the relative difference between $E_{\rm vis}^{\rm prompt}$
  of uniform positron events to
  that at the CD center shows a less than 0.05\% energy scale difference.}
  For the five realistic
  detector conditions above, with the {\it in situ} calibration, the
  bias can be \textcolor{black}{controlled} to below 0.3\%,
  as shown in
  table~\ref{summary_resolution}. Therefore, an additional 0.3\% systematic uncertainty 
  has been included in the positron energy scale in
  section~{\ref{sec:position_dep_bias}}. 

\subsection{Conclusion of the energy resolution}
The step-by-step downgrade of the non-uniformity calibration from the
ideal to the most realistic situation is summarized in
table~\ref{summary_resolution}. One sees that the constant term $b$ in
the energy resolution can be \textcolor{black}{optimized by} utilizing a single gamma source
deployed to about 250 points in a vertical plane of the CD, bootstrapped
to the entire CD using a smooth two-dimensional spline function in
$(r,\theta)$. This leads to an $\tilde a$ of 3.02\% for nominal JUNO detector, in
  agreement with the requirement put forward in Ref.~\cite{yellow-book}. 
  For detector imperfections (below the double-line 
  in table~\ref{summary_resolution}),
  the individual impact can be estimated by taking the difference of its $\tilde{a}$ and 3.02\% in quadrature. Although it is difficult to predict what may happen in a
  real detector, the individual imperfection can lead to a worst-case $\tilde{a}$ of 3.12\%, which is still sufficient to fulfill a $3\sigma$ determination of the MO~\cite{yellow-book}.
  Effects of combination of multiple imperfections are also studied in the simulation - approximately consistent with combining individual contributions in quadrature.  
  
\begin{table}[H]
  \centering
  \scriptsize
  \renewcommand\tabcolsep{3.0pt} 
  \begin{tabular*}{150mm}{@{\extracolsep{\fill}}ccccccc}
    \hline  
    Assumptions & $a$ & $b$ & $c$ & $\tilde{a}=\sqrt{a^{2} +(1.6b)^{2} + (\frac{c}{1.6})^{2}}$ & energy bias (\%)\\ 
    \hline
    Central IBDs & 2.62(2) & 0.73(1) & 1.38(4) & 2.99(1) & - \\
    Ideal correction & 2.57(2) & 0.73(1) & 1.25(4) & 2.93(1) & - \\
    Azimuthal symmetry & 2.57(2) & 0.78(1) & 1.26(4) & 2.96(1) & - \\
    Single gamma source & 2.57(2) & 0.80(1)& 1.24(4) & 2.98(1) & - \\
    Finite calibration points & 2.57(2) & 0.81(1)& 1.23(4) & 2.98(1) & - \\
    Vertex smearing(8 cm/$\sqrt{E(\rm MeV)}$) &2.60(2) & 0.82(1) & 1.27(4) & 3.01(1) & - \\\
    PMT QE random variations & 2.61(2) & 0.82(1) & 1.23(4) & 3.02(1) & 0.03(1)\\\hline
	\hline
    1\% PMT death (random)&2.62(2) & 0.84(1) & 1.23(5) & 3.04(1) & 0.09(1) \\
    1\% PMT death (asymmetric) &2.63(2) & 0.86(1) & 1.20(4) & 3.06(1) & 0.23(1) \\
    $Y_0$ reduced by 1\% & 2.62(2) & 0.85(1) & 1.25(4) & 3.05(1) & 0.09(1) \\
    $Y_0$ reduced by 5\% & 2.68(2) & 0.85(1) & 1.28(5) & 3.11(1) & 0.09(1) \\
    Absorption length reduced by 4\% &2.62(2) & 0.82(1) & 1.27(4) & 3.03(1) & 0.07(1)\\
    PMT single photon charge resolution (30\%) &2.72(2) & 0.83(1) & 1.23(5) & 3.12(1) & 0.08(1) \\
    \hline  
  	\end{tabular*}
    \caption{
	    \label{summary_resolution}
  		Energy resolution after sequential downgrade from the
    ideal to realistic calibration, considering all assumptions from {section~\ref{central_IBD}} to {section~\ref{sec:bias}}. 
	Values in parentheses indicate fitting uncertainties, and the uncertainty of $\tilde{a}$ has taken into account the correlations in $a$, $b$ and $c$.
	Each row from ``Azimuthal symmetry'' to ``PMT QE random variations'' indicates cumulative effects down to this row. 
	This gives an $\tilde a$ of 3.02\%  for nominal JUNO situation. 
	\textcolor{black}{Each line starting from ``1\% PMT death (random)'' represents an individual imperfection of the CD, which also includes effects up to the double-line (nominal $\tilde{a}$).}
	} 
\end{table}
\section{Conceptual design of the calibration system}
\label{sec:CalibSys}
\begin{figure}
  \centering
  \includegraphics[width=2.5in]{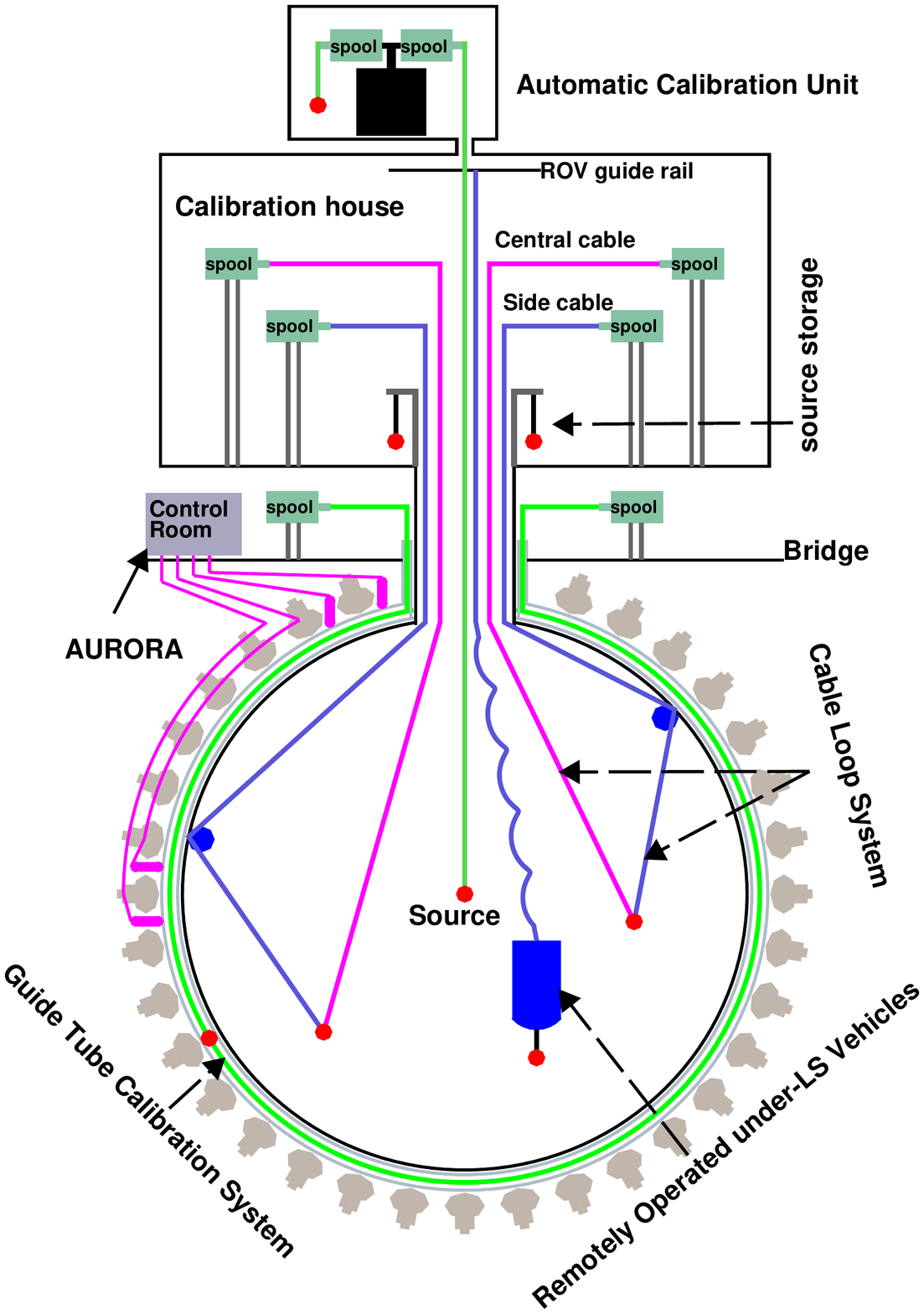}
  \caption{
  \label{overview_calibration_system}
  	Overview of the calibration system (not to scale),
    including the Automatic Calibration Unit (ACU), 
    two Cable Loop Systems (CLSs), the Guide Tube (GT), and the Remotely Operated Vehicle (ROV). 
    \textcolor{black}{The red points represent a source assembly described in figure~\ref{source_weight_QC}.}
    The AURORA is an
    auxiliary laser diode system to monitor the attenuation and
    scattering length of the LS, which is beyond the scope of this
    paper.}
\end{figure}

The hardware design of the calibration system is driven by
the strategy outlined in sections{~\ref{sec:scale_calib}}~and~{\ref{sec:resolution}}, 
to confront the challenge of energy scale and resolution calibration.
As demonstrated in
section~{\ref{sec:resolution}}, such a system should be capable to place 
a source along the central axis of the CD, on a circle at the
LS-acrylic boundary, and in the region in between. The corresponding
hardware design consists of several independent subsystems
(figure~\ref{overview_calibration_system}), which will be discussed in turn.
\textcolor{black}{To ensure the low background environment of the CD, the total requirement of introduced background from the calibration subsystems should be less than 0.5~Hz~\cite{juno-calib-background}.}

\subsection{Automatic Calibration Unit (ACU)}
The ACU is developed to do calibration along the central vertical axis $z$ of the
CD. The design is very similar to the ACU in the Daya Bay
experiment~\cite{dayabayACU}, with four independent spools mounted on
a turntable. Each spool is capable to unwind and deliver the source
via gravity through the central chimney of the CD, 
\textcolor{black}{with a better than 1~cm positioning precision in $z$.}
Three sources can be deployed regularly, including a neutron source (AmC), a gamma source
($^{40}$K), and a pulsed UV laser source carried by an optical fiber
with a diffuser ball attached to the end~\cite{zhangyuanyuanpaper}. To be flexible, the fourth
spool will carry a replaceable source, for example a radioactive
source or even a temperature sensor. Due to its simplicity and
robustness, we envision to use the ACU frequently during data taking to
\textcolor{black}{monitor the stability of} the energy scale, 
and to partially monitor the position non-uniformity.

\subsection{Guide Tube system (GT)}
The GT is a tube looped outside of the acrylic sphere along a
longitudinal circle \textcolor{black}{similar to that used in Double Chooz~\cite{DoubleChooz-Calib} 
and CUORE~\cite{CUORE-calib}.}
Within the tube, a radioactive source with cables
attached to both ends gets driven around with a positioning
precision of 3~cm. The design of this system is discussed in details in Ref.~\cite{GTCS}. Although physically outside of the CD, MeV-scale
gammas can easily penetrate the 12~cm acrylic and deposit energy into
the LS. 
The full absorption peak will be mixed with a leakage tail \textcolor{black}{due to energy deposition in the acrylic} 
that can be disentangled by fitting.
Based on the simulation studies in Ref.~\cite{GTCS}, 
this subsystem is sufficient to calibrate the CD non-uniformity at the boundary.
\subsection{Cable Loop System (CLS)}
As illustrated in figure~\ref{overview_calibration_system}, two CLSs
\textcolor{black}{will be installed} in the two opposite half-planes to deploy sources to
off-axis positions. \textcolor{black}{The design concept is inspired by 
those in SNO~\cite{SNO}, KamLAND~\cite{kamland-calib-paper}, and Borexino~\cite{Borexino-calib-paper}.}
For each CLS, two cables are attached to the
source, which also form a loop to deliver and retract the source. 
\textcolor{black}{Different sources can be interchanged on the CLS}.
The central cable goes upwards towards the north pole of the CD. The side
cable winds through an anchor on the inner surface of the acrylic
sphere, then also towards the north pole the CD. By adjusting the lengths of the two
cables, the source could be delivered ideally within an area bounded
by the vertical lines through the anchor and the central axis. More
realistic coverage measured by a CLS prototype is discussed in Ref.~\cite{zhangyuanyuan-new-paper}. 
The sources on the CLS will be positioned by an independent ultrasonic system 
to achieve a precision of 3~cm.

\subsection{Remotely Operated Vehicle (ROV)}
Locations other than the CLS plane could also turn out to be
important, 
\textcolor{black}{if significant local effects or azimuthal dependence were identified during data taking.}
These effects can be studied by physical
background events, for example, the spallation
neutrons. Alternatively, we also plan to have a ROV~\cite{ROV-paper} similar to that in
the SNO experiment~\cite{SNO-ROV}, \textcolor{black}{capable of deploying} a radioactive source
in \textcolor{black}{almost} the entire LS volume. As the CLS, the ROV also needs to work with the 
ultrasonic positioning system. The mechanical design also needs to be optimized
in size and surface reflection to minimize the loss of
photons. 
We envision that the ROV serves as a supplement to the ACU, CLS, and GT,
and should be deployed \textcolor{black}{infrequently}.

\subsection{Positioning system}
For all source deployment systems above, the control of source position is crucial. 
Table~\ref{position_precision} summarizes the expected precisions. 
The ACU and GT can achieve good positioning through accurate measurements of the cable lengths. For the
CLS, due to the self-weight of the cable and the friction in the loop,
cables do not run in straight lines, so calculations based on simple trigonometric relations introduce
significant uncertainties~\cite{zhangyuanyuan-new-paper}.  For the ROV, it is even more
complex as the positioning feedback is needed during the navigation. For these
purposes, an independent ultrasonic positioning system has been
developed~\cite{NWPU-paper}. Eight ultrasonic receivers will be mounted inside the
acrylic sphere, and the source deployed by the CLS or ROV will carry a
miniature ultrasonic emitter. Based on the prototype
tests, such a system is capable to provide a positioning precision to a level of 3 cm.

\begin{table}[H]
	\centering
	\renewcommand\tabcolsep{5.0pt}
    \begin{tabular*}{80mm}{@{\extracolsep{\fill}}cc}
          \hline  
		  System & Positioning precision [cm]\\
          \hline  
		  ACU & 1 \\
		  CLS & 3 \\
		  GT  & 3 \\
		  ROV & 3\\
          \hline  
    \end{tabular*}
  	\caption{\label{position_precision}Summary of expected positioning precision for each subsystem.}
\end{table}

\section{JUNO calibration program}
\label{sec:calib_program}
Based on all discussions above, we can now streamline the calibration
program. The rates of the radioactive source are set to be around 100
neutron or gamma emissions per second\footnote[4]{The only exception is $^{40}$K, which can only
be made with natural potassium salt with an approximate gamma rate of 1~Hz.}, so that the data rate during the
calibration does not differ too much from that during the neutrino
data taking ($\sim$1000 Hz). 

Similar to the Daya Bay calibration, we envision to separate the
program into comprehensive (but infrequent), weekly and monthly calibrations. 
As a requirement, source deployments should not introduce noticeable 
radioimpurity such as radon.
The nominal speed of the source movement is about 1 m/min. 

\subsection{Comprehensive calibration}
A comprehensive calibration is likely to be carried out at start
of the experiment to achieve a basic understanding of the CD
performance, and then a few times throughout the JUNO \textcolor{black}{life time}. 
Multiple sources will be deployed to the CD center to study the
non-linearity. 
The UV laser diffuser ball will be deployed to the CD
center and pulsed with a repetition rate of 50 Hz under eight different
intensities (equivalent energy from 0.3 MeV to 1~GeV), to allow
channel-wise instrumental calibration. 
In addition, the AmC neutron source will be deployed to
the full set of 250 points (figure~\ref{basic_coverage}) using the ACU,
CLS and GT. 
\textcolor{black}{ROV is envisioned not to be deployed at the beginning of the experiment, 
but when enough evidence of local effects or azimuthal dependence is accumulated.}
\textcolor{black}{An estimate time cost for a comprehensive calibration run is
shown in table~\ref{special_time}, with 100,000 events per calibration point at CD center (except $^{40}$K with 6000 events), 
including the travel time to move the source into designations. The total time of this campaign is expected to be about 48 hours.} 

\begin{table}[H]
	\centering
	\scriptsize
	\renewcommand\tabcolsep{3.5pt}
	\begin{tabular*}{150mm}{@{\extracolsep{\fill}}ccccccc}
          \hline  
          Source&Energy [MeV]&  Points & Travel time [min] & Data taking time [min]&Total time [min]\\ 
          \hline  
          Neutron (Am-C)  & 2.22  	   & 250&680 & 1262 & 1942\\
	  Neutron (Am-Be) & 4.4   	   & 1  & 58 & 17 & 75 \\
          Laser 	  & / 	  	   & 10  & 58 & 333 & 391 \\
          $^{68}$Ge 	  &$0.511 \times 2$& 1  & 58 & 17 & 75 \\
          $^{137}$Cs	  & 0.662 	   & 1  & 58 & 17 & 75 \\
          $^{54}$Mn 	  & 0.835 	   & 1  & 58 & 17 & 75 \\
          $^{60}$Co 	  & 1.17+1.33      & 1  & 58 & 17 & 75 \\
          $^{40}$K  	  & 1.461          & 1  & 58 &100 & 158 \\

          Total &/&/& 1086 & 1780 & 2866 ($\sim$ 48 h)\\
          \hline  
	\end{tabular*}
  	\caption{	
	\label{special_time}
	A baseline plan of a comprehensive calibration. 
  	\textcolor{black}{The AmC will be deployed into 250 
  	points utilizing the ACU, CLS and GT. 
  	All other sources will rely on the ACU only.
  	More sources at other locations are also possible.
  	}} 
\end{table}

\subsection{Weekly calibration}

\textcolor{black}{The weekly calibration is designed to track major 
changes of the detector properties such as variations}
in the light yield of the LS, PMT
gains, and electronics. As shown in table~\ref{weekly_time}, the
neutron source (AmC) will be deployed to five locations along the
central axis, each with one minute data taking, leading to a better-than 0.1\% 
statistical uncertainty to the gamma peak.
The UV laser intensity scan will also 
be taken at the CD center. The total time of weekly calibration is about 2.4 hours.
\begin{table}[H]
	\scriptsize
	\renewcommand\tabcolsep{4.0pt} 
	\begin{tabular*}{150mm}{@{\extracolsep{\fill}}ccccccc}
          \hline  
          Source&Energy [MeV]& Points & Travel time [min] & Data taking time [min]& Total time [min]\\ 
          \hline  
          Neutron (Am-C) & 2.22 & 5 & 58 & 5  & 63\\
	  Laser 	 & /    & 10& 58 & 20 & 78\\
          Total 	 & /    & / & 116& 25 & 141 ($\sim$2.4 h)\\
          \hline  
	\end{tabular*}
  	\caption{\label{weekly_time}Weekly calibration \textcolor{black}{using the ACU.}}
\end{table}

\subsection{Monthly calibration} 
\label{sec:monthly_calib}
\textcolor{black}{The monthly calibration goes through a limited number of positions}
in figure~\ref{basic_coverage}, 
\textcolor{black}{given that a comprehensive calibration has been performed at the start of the experiment, and that}
\textcolor{black}{the temporal variations such as the PMTs and optical properties of the LS are minimal.}
As shown in
table~\ref{monthly_time}, the ACU, CLS, and GT will all be operated
during the calibration. 
The UV laser will also be deployed to the same 
locations as the AmC along the central axis. To
balance the extensiveness and time cost, we select the full sets of 27
and 23 points from the ACU and GT, respectively, and 40 representative points in CLS
to monitor the non-uniformity of the CD. The total time for a monthly calibration is expected to be 11.2 hours.


\begin{table}[H]
	\scriptsize
	\renewcommand\tabcolsep{4.0pt} 
	\begin{tabular*}{150mm}{@{\extracolsep{\fill}}cccccccc}
                  \hline  
                  System&Source& Points & Travel time [min] & Data taking time [min]&Total time [min]\\ 
                  \hline  
                  ACU  &Neutron (Am-C) &   27 & 93 & 27 & 120\\
                  ACU  &Laser 	       &   27 & 93 & 54 & 147\\
                  CLS  &Neutron (Am-C) &   40 & 293& 40 & 333\\
                  GT   &Neutron (Am-C) &   23 & 50 & 23 & 73\\
		  Total&/	       &   /  & 529&144 & 673 ($\sim$11.2 h)\\
                  \hline  
	\end{tabular*}
  	\caption{\label{monthly_time}Monthly calibration \textcolor{black}{with ACU, CLS and GT.}}
\end{table}


\section{Summary}
\label{sec:summary}
We have carried out a comprehensive study to develop a multi-faceted
calibration strategy to secure JUNO's full potential 
in the determination of
the neutrino mass ordering. This study is based on the most
up-to-date JUNO simulation software, including all major features \textcolor{black}{of}
the detector design. We demonstrate that using various gamma and neutron sources, 
cosmogenic $^{12}$B, in combination with a pulsed UV laser, 
the nonlinear energy scale of
the positrons can be determined to a sub-percent level within the entire energy range of
the IBDs. 
The novel dual calorimetry allows the clean determination of
instrumental non-linearity, leading to robust and independent control of other non-linearity
and non-uniformity effects.
We also have developed a multi-positional source
deployment strategy to optimize the energy resolution. With a selection of 250 key
positions in a vertical plane of the detector and by utilizing the
azimuthal symmetry, an effective energy resolution $\tilde a$ of 3.02\% is achieved with the
nominal JUNO detector parameters.
This calibration plan requires a multi-component source deployment hardware
including a vertical spooling system covering the central axis of the
CD, a guide tube system attached to the acrylic sphere, two cable
loops to cover a large fraction of area in a vertical plane, and a
supplementary Remotely Operated Vehicle. In the end, we separate
the calibration tasks into different frequency categories to ensure
both the timeliness and the comprehensiveness. \textcolor{black}{Approximately 3\% of the total live time is dedicated to calibration}. We demonstrate that
with such a calibration strategy, JUNO's challenging requirements on the
neutrino energy spectrum measurement can be achieved.


%
%
%



\begin{thebibliography}{99}

%
%



\bibitem{PDG}M.~Tanabashi \textit{et al.} (Particle Data Group), \emph{Review of Particle Physics}, \emph{Phys. Rev. D} {\bf 98} (2018) 030001.
\bibitem{yellow-book}Fengpeng An et al, \emph{Neutrino physics with JUNO}, \emph{Phys. G: Nucl. Part. Phys.} {\bf 43} (2016) 030401.
\bibitem{petcov-original-paper}S. T. Petcov and M. Piai, \emph{The LMA MSW solution of the solar neutrino problem, inverted neutrino mass hierarchy and reactor neutrino experiments}, \emph{Phys. Lett. B} {\bf 533} (2002) 94.
\bibitem{petcov2}S. Choubey, S. T. Petcov and M. Piai, \emph{Precision neutrino oscillation physics with an intermediate baseline reactor neutrino experiment}, \emph{Phys. Rev. D} {\bf 68} (2003) 113006.
\bibitem{learned}J. Learned, S. Dye, S. Pakvasa and R. Svoboda, \emph{Determination of neutrino mass hierarchy and $\theta_{13}$ with a remote detector of reactor antineutrinos}, \emph{Phys. Rev. D} {\bf 78} (2008) 071302.
\bibitem{zhanliang-original-paper-2008}L. Zhan, Y. Wang, J. Cao and L. Wen, \emph{Determination of the neutrino mass hierarchy at an intermediate baseline}, \emph{Phys. Rev. D} {\bf 78} (2008) 111103.
\bibitem{zhanliang-original-paper-2009}L. Zhan, Y. Wang, J. Cao and L. Wen, \emph{Experimental requirements to determine the neutrino mass hierarchy using reactor neutrinos}, \emph{Phys. Rev. D} {\bf 79} (2009) 073007.
\bibitem{zhanliang-original-paper-2013}Li Y F, Cao J, Wang Y and Zhan L, \emph{Unambiguous determination of the neutrino mass hierarchy using reactor neutrinos}, \emph{Phys. Rev. D} {\bf 88} (2013) 013008.
\bibitem{JUNOCDR}Zelimir Djurcic et al. (JUNO Collaboration), \emph{JUNO Conceptual Design Report}, arxiv:1508.07166.
\bibitem{Geant4}Agostinelli S et al. (GEANT4), \emph{GEANT4--a simulation toolkit}, \emph{Nucl. Instrum. Meth. A} {\bf 506} (2003) 250-303.
\bibitem{SNIPER}Lin, Tao et al., \emph{The Application of SNiPER to the JUNO Simulation}, \emph{J. Phys. Conf. Ser.} {\bf 898} (2017) 042029.
\bibitem{JUNOLSAbsL}Y. Zhang et al., \emph{A complete optical model for liquid-scintillator detectors}, \emph{Nucl. Instrum. Meth. A} {\bf 967} (2020) 163860.
\bibitem{YuZeyuan_LSpaper}A. Abusleme at al. (JUNO Collaboration), \emph{Optimization of the JUNO liquid scintillator composition using a Daya Bay antineutrino detector}, arxiv:2007.00314. 
\bibitem{Zhouxiang_Paper}Zhou X et al., \emph{Rayleigh scattering of linear alkylbenzene in large liquid scintillator detectors}, \emph{Rev Sci Instrum.} {\bf 86(7)} (2015) 073310.
\bibitem{geant4-physics}Geant4 Collaboration, \emph{Geant4 Physics reference Manual Release 10.5}, (2019)
\bibitem{borexino-positronium-paper}D. Franco, G. Consolati and D. Trezzi, \emph{Positronium signature in organic liquid scintillators for neutrino experiments}, \emph{Phys. Rev. C} {\bf 83} (2011) 015504.
\bibitem{Ortho-positronium-double-chooz}Abe, Y., dos Anjos, J.C., Barriere, J.C. et al., \emph{Ortho-positronium observation in the Double Chooz experiment}, \emph{J. High Energ. Phys.} {\bf 2014} (2014) 32.
\bibitem{measure-lifetime-ortho}Mario Schwarz et al., \emph{Measurements of the lifetime of orthopositronium in the LAB-based liquid scintillator of JUNO}, \emph{Nucl. Instrum. Meth. A} {\bf 922} (2019) 64-70.
\bibitem{kamland-calibration-paper}J. Detwiler, \emph{MEASUREMENT OF NEUTRINO OSCILLATION WITH KAMLAND}, \emph{Standford University} (2005).
\bibitem{dayabay-first-shape-paper}F. P. An et al. (Daya Bay Collaboration), \emph{Spectral Measurement of Electron Antineutrino Oscillation Amplitude and Frequency at Daya Bay}, \emph{Phys. Rev. Lett.} {\bf 112} (2014) 061801.
\bibitem{semi-ana}P. Kampmann and Y. Cheng and L. Ludhova, \emph{A semi-analytical energy response model for low-energy events in {JUNO}}, \emph{JINST} {\bf 15} (2020) P10007--P10007.
\bibitem{dayabay-calib-paper}D. Adey et al. (Daya Bay Collaboration), \emph{A high precision calibration of the nonlinear energy response at Daya Bay}, \emph{Nucl. Instrum. Meth. A} {\bf 940}(2019) 230-242.
\bibitem{kamland-cosmogenic-bkg-paper}S. Abe et al. (KamLAND Collaboration), \emph{Production of radioactive isotopes through cosmic muon spallation in KamLAND}, \emph{Phys. Rev. C} {\bf 81} (2010) 025807.
\bibitem{borexino-cosmogenic-bkg-paper}G Bellini et al. (Borexino collaboration), \emph{Cosmogenic Backgrounds in Borexino at 3800 m water-equivalent depth}, \emph{Journal of Cosmology and Astroparticle Physics} {\bf 2013} (2013) 049.
\bibitem{zhangyuanyuanpaper}Y. Zhang et al., \emph{Laser calibration system in JUNO}, \emph{JINST} {\bf 14} (2019) P01009.
\bibitem{kamland-calib-paper}B E Berger el al. (KamLAND Collaboration), \emph{The KamLAND full-volume calibration system}, \emph{JINST} {\bf 4} (2009) P04017.
\bibitem{Borexino-calib-paper}B.~Caccianiga and A.~C.~Re, \emph{The calibration system for the Borexino experiment}, \emph{Int. J. Mod. Phys. A} {\bf 29} (2014) 1442010.
\bibitem{DoubleChooz-Calib}Igor Ostrovskiy et al., \emph{Double Chooz Calibration}, \emph{Nuclear Physics B (Proc. Suppl.)} {\bf 229-232} (2012) 431.
\bibitem{private_comm1}Feiyang Zhang, private communication.
\bibitem{fitmethod}Jia-Hua Cheng et al., \emph{Determination of the total absorption peak in an electromagnetic calorimeter}, \emph{Nucl. Instrum. Meth. A} {\bf 827} (2016) 165-170.
\bibitem{Double-Chooz-nature-physics}H. de Kerret et al. (Double Chooz Collaboration), \emph{Double Chooz $\theta_{13}$ measurement via total neutron capture detection}, \emph{Nat. Phys.} {\bf 16} (2020) 558-564.
\bibitem{zhangyuanyuan-new-paper}Y. Zhang et al., \emph{Cable Loop Calibration System for Jiangmen Underground Neutrino Observatory}, arxiv:2011.02183.
\bibitem{vertex-rec}Q. Liu et al., \emph{A vertex reconstruction algorithm in the central detector of JUNO}, \emph{JINST} {\bf 13} (2018) T09005.
\bibitem{man_in_prep}Manuscript in preparation.
\bibitem{Lux-single-photons}D. S. Akerib et al. (LUX Collaboration), \emph{Extending light WIMP searches to single scintillation photons in LUX}, \emph{Phys. Rev. D} {\bf 101} (2020) 042001.
\bibitem{huangyongbo-papper}Yongbo Huang et al., \emph{The Flash ADC system and PMT waveform reconstruction for the Daya Bay Experiment}, \emph{Nucl. Instrum. Meth. A} {\bf 895} (2018) 48-55.
\bibitem{juno-calib-background}Feiyang Zhang et al., \emph{The background of the JUNO calibration system}, manuscript in preparation.
\bibitem{dayabayACU}J. Liu et al., \emph{Neutron Calibration Sources in the Daya Bay Experiment}, \emph{Nucl. Instrum. Meth. A} {\bf 750} (2014) 19.
\bibitem{CUORE-calib}Jeremy S. Cushman et al., \emph{First results from the CUORE experiment}, \emph{Nucl. Instrum. Meth. A} {\bf 844} (2017) 32-44.
\bibitem{GTCS}Yuhang Guo et al., \emph{Design of the Guide Tube Calibration System for the JUNO experiment}, \emph{JINST} {\bf 14} (2019) T09005.
\bibitem{SNO}J. Boger et al., \emph{The Sudbury neutrino observatory}, \emph{Nucl. Instrum. Meth. A} {\bf 449} (2000) 172-207.
\bibitem{ROV-paper}K. Feng et al., \emph{A novel remotely operated vehicle as the calibration system in JUNO}, \emph{JINST} {\bf 13} (2018) T12001.
\bibitem{SNO-ROV}J. F. Amsbaugh et al., \emph{An array of low-background $^{3}$He proportional counters for the Sudbury Neutrino Observatory}, \emph{Nucl. Instrum. Meth. A} {\bf 579} (2007) 1054-1080.
\bibitem{NWPU-paper}G. L. Zhu, J. L. Liu, Q.~Wang, M. J. Xiao and T. Zhang, \emph{Ultrasonic positioning system for the calibration of central detector}, \emph{Nucl. Sci. Tech.} {\bf 30} (2019) 5.




\end{thebibliography}
\end{document}